\documentclass[pre,a4paper,twocolumn,superscriptaddress,%
floatfix]{revtex4}

\usepackage[T1]{fontenc}
\usepackage[utf8]{inputenc}

\usepackage{graphicx}
\usepackage{color}
\usepackage{amsmath}
\usepackage{csquotes}
\usepackage{braket}
\usepackage{bm}
\usepackage[normalem]{ulem}

\begin{document}

\title{Nonequilibrium distribution for stochastic thermodynamics}

\author{Jean-Luc Garden}
\email{jean-luc.garden@neel.cnrs.fr}
\affiliation{Univ. Grenoble Alpes, CNRS, Grenoble INP, Institut N\'{E}EL,
38000 Grenoble, France}

\date{\today}

\begin{abstract}
We extend the canonical Gibbs distribution, originally formulated for systems at equilibrium, to systems driven out of equilibrium. The stochastic dynamics of a small system are described by a probability distribution over discrete energy levels. Within this framework, we derive a microscopic expression for work and introduce a microscopic definition of entropy production (defined here in terms of the uncompensated heat of Clausius) during a nonequilibrium stochastic process. Work and entropy production share a common origin arising from variations of the system energy. The proposed framework allows us to recover the nonequilibrium work relation and to establish an equivalent identity for the heat exchanged during a work protocol. Finally, we show that the fluctuations of work and heat governed by the extended canonical distribution follow directly from the fluctuation theorem for entropy production.
\end{abstract}

\maketitle

\section{Introduction}  
Stochastic thermodynamics investigates the thermodynamic behavior of small systems. These systems are sufficiently small for their properties to be significantly influenced by interactions with reservoirs. At the same time, they remain large enough for their thermodynamic states to be described by a limited set of macroscopic, measurable variables. However, due to the discrete nature of matter and the effects of thermal fluctuations at microscopic scales, the fluctuations of these macroscopic variables become non-negligible in small systems. 
In recent decades, a series of theoretical results known as fluctuation theorems, supported by high-precision experiments, have been developed at the mesoscopic scale. For a comprehensive overview of stochastic thermodynamics, we refer the reader to the review articles and book \cite{EVANS1, MARCONI, JAR1, SEIFERT, SEKI1}.
At this scale, measurable thermodynamic quantities (such as the work performed during a process) become random variables characterized by probability distributions. This implies that repeated measurements under identical experimental conditions yield different outcomes, statistically distributed according to a well-defined probability law. Importantly, this probability distribution depends on the rate at which external control parameters are varied, such as those used to exchange work between the system and its environment.
Therefore, in stochastic thermodynamics, it is not only essential for the system to be small, but also for its time evolution to occur out of equilibrium for at least a part of the process. This nonequilibrium condition is a fundamental prerequisite in stochastic thermodynamics. This implies that, during the evolution of the system, the condition of statistical equilibrium no longer holds. In the absence of statistical equilibrium, density in phase of the system within the considered statistical ensemble and the density probability in the given element of volume in the phase space are no longer constants of motion \cite{GIBBS}. Under such conditions, nonconservative forces must be considered \cite{GIBBS}. These forces generally depend not only on the generalized coordinates $q_{i}$ and the external control parameters $a_{i}$ (which represent the influence of external bodies) but also on the generalized momenta $p_{i}$ \cite{GIBBS}.
To describe a system at the macroscopic level, only a few macroscopic variables are typically required. At equilibrium or during an equilibrium transformation, the internal variables characterizing the state of the system depend solely on the temperature and the external control parameters \cite{BAZA}. For instance, the internal energy $U$ of the system depends on the temperature and an external parameter $\lambda$. The parameter $\lambda$ represents the extensive variable associated with work exchange between the system and a work reservoir (e.g., $\lambda$ corresponds to the vessel volume controlled by the piston in the case of mechanical work). In contrast, outside of equilibrium, additional variables (denoted $\xi_{i}$) are necessary to describe the instantaneous state of the system \cite{BAZA, PRIGO, DEDONDER}. These variables are microscopically related to the aforementioned nonconservative forces. The set $\xi_{i}(t)$ captures the time-dependent evolution of the internal state of the system when both the temperature and $\lambda$ are held constant.

\textit{Basic idea}\textemdash Our work is based on two central ideas. First, since stochastic thermodynamics lies at the interface between statistical physics and thermodynamics, an appropriate statistical averaging of the relevant microscopic random quantities must reproduce the laws of thermodynamics. Indeed, fluctuation theorems and nonequilibrium relations (such as the nonequilibrium work relation or the Jarzynski equality) involve experimentally measurable macroscopic quantities, including the performed work or the free-energy difference, which are thermodynamic observables. In other words, the statistical average of fluctuating microscopic quantities, computed using a suitable probability distribution, must yield the familiar macroscopic thermodynamic quantities. However, a more stringent requirement arises: Because the system under study is transiently out of equilibrium, the statistical averaging should specifically recover the established laws of macroscopic nonequilibrium thermodynamics.
Second, it is well known that a macroscopic system, classically described by a Hamiltonian $\mathcal{H}(q_{i},p_{i})$ and a continuous distribution function $\rho (q_{i},p_{i})$, can also be represented (if sufficiently small) in terms of a discrete set of energy levels $E_{i}$ with an associated discrete probability distribution $P_{i}$. Each energy level $E_{i}$ corresponds to a Hamiltonian $\mathcal{H}(q_{i},p_{i})$ in the statistical ensemble. The microscopic state labeled by $i$ thus corresponds, in the limit of a continuous probability density, to an infinitesimal phase-space element of volume $\delta q_{i} \delta p_{i}$. The ensemble we consider is the canonical ensemble, i.e., the set of all possible values of $E_{i}$, as formulated by Gibbs \cite{GIBBS}. This framework is appropriate when fluctuations arise due to coupling with thermal or work reservoirs. The statistical average of a given random quantity over the canonical ensemble corresponds to a sum over all number $i$ with $1 \le i \le N$ (more precisely over all the $N$ possible states occupied by the system), weighted by their respective probabilities of occurrence. However, the probability distribution must be modified to reflect the nonequilibrium nature of the process. To this end, we propose a straightforward extension of the Gibbs canonical distribution to include the internal variables $\xi $. This extension naturally introduces a time dependence into the relevant quantities, even when both the temperature and the work-related parameter $\lambda$ are held constant. A justification for the existence and validity of this nonequilibrium distribution is provided in Appendix B.

The structure of the paper is as follows. In the first part, we summarize the laws of macroscopic thermodynamics for systems out of equilibrium. This section builds on the foundational work of De Donder, Prigogine, Defay, and others from the Belgian school of thermodynamics \cite{PRIGO, DEDONDER}. This section clarifies the role of internal variables relative to external ones and their relation to entropy production in our framework.  All thermodynamic quantities are understood as statistical averages over the accessible states of the system within the statistical ensemble (for clarity, the overlines denoting statistical averages are omitted in this section). In the second part, we introduce a nonequilibrium probability distribution for a multilevel system, representing an extension of the canonical Gibbs distribution used for systems at equilibrium. Here, due to the small size of the system and the nonequilibrium nature of the process, both energy and entropy are random variables that depend on an additional macroscopic variable $\xi(t)$. Consequently, each realization $E_i$ and $S_i$ obtained in a stochastic experiment, which generally depends on $\lambda(t)$, also depends on $\xi(t)$. By taking infinitesimal changes of this extended canonical distribution with respect to the macroscopic variables of the system, we recover the classical expression for work as defined in stochastic thermodynamics. Furthermore, we derive a quantity which we refer to as the \textit{the random uncompensated heat of Clausius}. This quantity is directly related to the concept of entropy production at the microscopic scale, and within our approach, entropy production acquires a clear microscopic interpretation. While defining heat at the microscopic level is more subtle than defining work, we treat heat as a random variable as well, inherently linked to the stochastic nature of work and entropy production during a process. Using the extended canonical distribution, we perform statistical averages that allow us to recover the macroscopic laws of nonequilibrium thermodynamics for all the relevant quantities. This result reinforces the consistency and validity of our approach.
In the next step, we rederive the nonequilibrium work relation and obtain an analogous expression for the random heat exchanged during the process. We then derive a nonequilibrium identity, closely related to the nonequilibrium work relation, both following solely from the fluctuation theorem for entropy production.

\section{Macroscopic nonequilibrium thermodynamics} 
This section provides a brief overview of classical macroscopic nonequilibrium thermodynamics. We present two equivalent formulations of the uncompensated heat of Clausius. These expressions can be found in the foundational works of De Donder and his school \cite{PRIGO, DEDONDER}. We consider, as illustrated in Fig.~1, a textbook thermodynamic system consisting of a gas of atoms or molecules confined in a variable volume controlled by a movable piston. 
\begin{figure}[!htbp]
  \centering
  \includegraphics[width=6 cm]{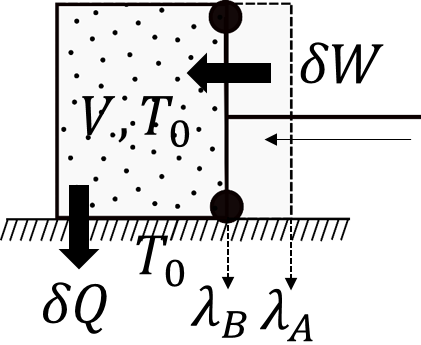}
  \caption{Schematic representation of a thermodynamic system consisting of particles (atoms or molecules) confined in a vessel in perfect thermal contact with a heat bath that maintains a constant temperature $T_{0}$. The system is also coupled to a work reservoir through an externally controlled vessel volume with a movable piston, which can perform or extract mechanical work. When work is supplied to the system through a change in the external parameter $\lambda=V_{v}$ (the vessel volume), heat is transferred isothermally to the bath.}
  \label{fig:Fig1}
\end{figure}
This volume is assumed to be in perfect thermal contact with a heat reservoir at constant temperature $T_{0}$. The system is also coupled to a work reservoir characterized by a generalized force $f$ (pressure, defined here as force per unit area), conjugate to an externally controllable parameter $\lambda $. The internal energy $U$ is taken to be a state function of the system, and its infinitesimal change arises solely from energy exchanges with the thermal and mechanical surroundings. This leads to the first law of thermodynamics for a closed system, where no exchange of matter with the environment occurs
\begin{equation}
\label{eq:FirstLaw1}
  dU= \delta Q + \delta W  .
\end{equation}
The quantity $\delta Q$ denotes the infinitesimal heat exchanged between the system and the thermal reservoir, while $\delta W$ represents the infinitesimal work associated with variations in the external control parameter. These quantities are not exact differentials, in contrast with the total differential $dU$, which characterizes the change in internal energy, a state function. By defining work solely in terms of changes in the external parameter $\lambda $, the first law of thermodynamics enables a definition of heat as the difference between the change in internal energy and the work performed on (or by) the system. At the macroscopic level, the infinitesimal work exchange is consequently defined as
\begin{equation}
\label{eq:dW1}
  \delta W=-fd\lambda .
\end{equation}
Here, $d\lambda $ denotes the variation of an external control parameter (i.e., the volume of the vessel), and $f$ is the corresponding intensive mechanical force (e.g., the external pressure) conjugate to this parameter. There exist as many external parameters as there are distinct ways to perform work on the system. The equilibrium thermodynamic state of the system is characterized by the temperature $T$ and the external parameter $\lambda $. In this framework, the mechanical force $f$ depends on both $T$ and $\lambda $, such that
\begin{equation}
\label{eq:f}
  f=f(T,\lambda)=-\left[\partial F(T,\lambda)/\partial \lambda \right]_{T} .
\end{equation}
The function $F$ is the Helmholtz free energy, defined by $F=U-TS$, where $S$ denotes the entropy of the system. Like $U$, both $F$ and $S$ are thermodynamic state functions. 

For a system out of equilibrium, the second law of thermodynamics can be expressed in the form of an equality  \cite{PRIGO, DEDONDER}
\begin{equation}
\label{eq:SecondLaw1}
  dS=d_{e}S+d_{i}S .
\end{equation}
The term $d_{e}S=\delta Q/T$ represents the entropy change associated with the exchange of heat between the system and the thermal bath at temperature $T$. This provides an alternative expression for the heat transfer, formulated in terms of entropy rather than internal energy, as in the first law. The term $d_{i}S\geq\ 0$ denotes the positive entropy production arising from irreversible processes occurring within the system during a transformation. It is explicitly expressed as
\begin{equation}
\label{eq:diS1}
  d_{i}S=\frac{A}{T}d\xi .
\end{equation}
The variable $\xi$ is an internal state parameter that quantifies the progress of the irreversible process within the system. The quantity $A$ is the thermodynamic affinity associated with this process. It defines a new state function $A(T,\lambda , \xi)$. Out of equilibrium, all thermodynamic state functions become functions of the three independent variables $T$, $\lambda$, and $\xi$. Their infinitesimal variations remain exact differentials but now include a third contribution absent in equilibrium. The physical meanings of the internal variable $\xi$, the affinity $A$, and the entropy production $d_{i}S$ are discussed in Appendix A for the model system illustrated in Fig.~1. Their microscopic counterparts have also been examined in the stochastic thermodynamics literature for this model system. The entropy production and uncompensated heat of Clausius are related through the following expression
\begin{equation}
\label{eq:diS2}
 d_{i}S=\delta Q'/T. 
\end{equation}
With these definitions, it follows that at constant temperature, the product $TS$ is itself a thermodynamic state function. In contrast with Eq.~(\ref{eq:FirstLaw1}), the heat exchange can now be expressed as the infinitesimal variation of this state function, minus the uncompensated heat of Clausius
\begin{equation}
\label{eq:heat2}
 \delta Q=d(TS)-\delta Q'.
\end{equation}
The quantity $\delta Q'$ should be interpreted as the amount of heat generated internally within the system by irreversible relaxation processes that have not yet had time to be transferred to the thermal reservoir during the time interval $dt$. This delay arises from the finite timescales associated with relaxation mechanisms intrinsic to the ongoing irreversible processes. It is crucial in our approach to emphasize that the exchanged heat $\delta Q$ and the uncompensated heat $\delta Q'$ are of fundamentally different physical origin. Both correspond to entropy variations: The former represents entropy exchange, whereas the latter corresponds to internal entropy production. This distinction is central to our forthcoming developments, wherein we demonstrate that, at the microscopic level, the uncompensated heat has the same origin as the work exchanged with external bodies and is therefore related to the variation of the system energy (or of the Hamiltonian in the continuous case). In the general (nonisothermal) case, the second law expressed in Eq.~(\ref{eq:heat2}) must therefore be reformulated to highlight the role of uncompensated heat
\begin{equation}
\label{eq:dQ'1}
  \delta Q'=d(TS)-\delta Q-SdT\geq\ 0.
\end{equation}
By substituting this last expression for the heat exchange into Eq.~(\ref{eq:FirstLaw1}) and employing the definition of the Helmholtz free energy $F$, we obtain
\begin{equation}
\label{eq:dQ'2}
  \delta Q'=\delta W-dF-SdT\geq\ 0.
\end{equation}
The two preceding expressions are fully equivalent, representing alternative formulations of the second law of thermodynamics. The uncompensated heat of Clausius is expressed in terms of different thermodynamic quantities in each case: The first relation, Eq.~(\ref{eq:dQ'1}), involves heat and entropy, while the second, Eq.~(\ref{eq:dQ'2}), is formulated in terms of work and free energy. At this point, it is reasonable to conjecture that if the nonequilibrium work relation in stochastic thermodynamics establishes a connection between fluctuating work and the free-energy difference [as in Eq.~(\ref{eq:dQ'2})], then a corresponding relation might exist linking random heat exchange to the entropy change [as in Eq.~(\ref{eq:dQ'1})]. Our first objective is to recover both nonequilibrium thermodynamic expressions, Eqs.~(\ref{eq:dQ'1}) and~(\ref{eq:dQ'2}), through direct averaging of the corresponding microscopic quantities, as developed in the next section within the framework of nonequilibrium statistical physics. The second objective is to derive the nonequilibrium work relation, along with a new, equivalent expression for the heat exchange with the thermal reservoir, based on a statistical description that incorporates irreversible processes.

\section{Nonequilibrium statistical physics}
\subsection{Classical statistical mechanics}
The system illustrated in Fig.~1 consists of a large number of particles characterized by generalized positions $q_{i}$ and generalized momenta  $p_{i}$ ($1 \le i \le n$). Its total energy is described by the Hamiltonian $\mathcal{H}(q_{i},p_{i})$, a function of the generalized coordinates that governs the dynamics through Hamilton’s equations of motion. Following Gibbs, the evolution of a system with $n$ degrees of freedom can be effectively described in terms of a statistical ensemble composed of $N$ identical replicas, each governed by the same Hamiltonian $\mathcal{H}$, but differing in the probability to find these systems into a delimited volume in the phase space \cite{GIBBS}. To account for interactions with the environment and energy exchange with external bodies, the Hamiltonian is further extended to depend on generalized external coordinates $a_{i}$, through its potential energy term  $\epsilon _{q}$ \cite{GIBBS}. Statistical equilibrium is defined by the constancy of the density in phase or by the constancy of the volume in the phase space over time. That is, the probability that a system taken at random from an ensemble canonically distributed and falling within any given limits of phase is constant. This condition of statistical equilibrium holds only if all forces acting on the system are conservative \cite{GIBBS}. In such cases, the forces derive from a potential, and the work performed on (or by) external bodies corresponds to an exact differential of the potential energy. In contrast, the presence of nonconservative forces renders work not an exact differential. Such forces lead to macroscopic manifestations of dissipation, as included in the Rayleigh dissipation function \cite{RAYLEIGH}. At the macroscopic scale, dissipation is associated with irreversible entropy production, corresponding to entropy that does not have sufficient time to be exchanged with the surroundings through heat transfer. Here, we introduce the effect of nonconservative microscopic interactions through additional generalized internal coordinates $\zeta_{i}$, in analogy with the external coordinates $a_{i}$. These new variables imply that the forces now also depend on the generalized momenta. The Hamiltonian is thus extended to $\mathcal{H}(q_{i},p_{i},a_{i},\zeta_{i})$. As a consequence, the thermodynamic process acquires an explicit dependence on the rate at which the external parameters are varied. In what follows, we show how the canonical Gibbs distribution must be naturally generalized to incorporate these non-conservative effects.

\subsection{Extended Gibbs canonical distribution}
Here, $X$ denotes a random variable of a stochastic process, whereas $X_{i}$ refers to a specific realization obtained when a state of number $i$ is sampled (i.e., in a single experimental run). Accordingly, the thermodynamic variables in the previous Sec.~II are implicitly understood as statistical averages over all the states, weighted by the probabilities $P_{i}$ ($1 \le i \le N$).

The macroscopic system illustrated in Fig.~1 is now rescaled to represent a sufficiently small system that can be modeled as a microscopic system with $N$ discrete energy levels $E_{i}$. The occupation probability of each level is $P_{i}$. To each energy level $E_{i}$ associated with microstate $i$ corresponds a Hamiltonian $\mathcal{H}(q_{i},p_{i},a_{i},\zeta_{i})$, defined over a small element of phase–space volume $\delta q_{i} \delta p_{i}$ (or small element of extension in phase) \cite{GIBBS}. In the absence of the nonconservative coordinates $\zeta_{i}$, the system is in statistical equilibrium. In this case, the index of probability of the phase (or the logarithm of the probability), which is a linear function of energy is written \cite{GIBBS}
\begin{equation}
\label{eq:Eta}
  \eta =\log P=\frac{\Psi-\epsilon}{\Theta } .
\end{equation}
This is the canonical distribution originally introduced by Gibbs \cite{GIBBS}. Let $\epsilon$ denote the energy and $\Theta>0$ the modulus of the distribution. $\Psi$ denotes the energy value for which the probability density $P$ equals unity. The function introduced above acts as a normalization factor, ensuring that the probabilities over all stochastic events sum to unity. The main argument used by Gibbs was that the index of probability is a decreasing linear function of the energy, as given in Eq.~(\ref{eq:Eta}). The normalization condition above, i.e., the sum over all states, excludes laws of the type $P=\epsilon \times \text{constant}$ or $P=\text{constant}$. In the presence of nonconservative coordinates $\zeta_{i}$, the system is no longer in statistical equilibrium. In this case the parameter $\eta$, which characterizes the index of probability, is no longer constant in time for evolving systems of the ensemble. During a process in which the density in phase of the statistical ensemble is displaced (such as when external bodies perform work on the system) the probability density and the index of probability evolve in time with the system energy. In general, the time-dependent probabilities $P_{i}$ are governed by linear first-order differential equations (master equations) involving probability transition rates between energy levels \cite{JAR3}. Here, to account for the time evolution of the probabilities, we propose to incorporate it directly into the canonical form of the distribution through the internal parameter $\xi$, which couples to the internal coordinates $\zeta_{i}$ in the Hamiltonian. The underlying assumption is that the microscopic degrees of freedom remain in thermal equilibrium with the reservoir for fixed instantaneous values of $(\lambda,\xi)$, whereas the internal variable $\xi$ may deviate from its equilibrium value and evolve according to its own kinetics
\begin{equation}
\label{eq:Pi0}
  P=e^{\eta}=\frac{e^{-\beta E(\lambda,\xi)}}{Z} .
\end{equation}
In Appendix B, we derive this expression by considering the system of interest as a small subsystem of a large, isolated system with fixed energy, described by the microcanonical ensemble. The partition function or sum over all states is given by
\begin{equation}
\label{eq:Z}
  Z(\beta, \lambda,\xi)=e^{-\beta \Psi(\beta, \lambda,\xi)}=\sum_{i=1}^{N}e^{-\beta E_{i}(\lambda,\xi)} .
\end{equation}
In contrast with the canonical ensemble, the probability $P$ (and thus the coarse-grained statistical entropy $S=-k_{B}\eta$) as well as $Z$ now depend not only on the inverse of the distribution modulus $\beta=1/\Theta$ and on the external control parameter $\lambda$, which couples to the coordinates $a_{i}$, but also on an internal parameter $\xi$, which couples to the coordinates $\zeta_{i}$. Accordingly, the energy $E$ (and, by assumption, all levels $E_{i}$) are functions of both $\lambda$ and $\xi$ but not of $\beta$. For simplicity, we restrict our analysis to a single external parameter $\lambda$, which governs the exchange of work, and a single nonequilibrium parameter $\xi$. However, in more complex systems subject to multiple external driving forces (mechanical, electrical, magnetic, etc.), additional control parameters $\lambda_{j}$ may be introduced, each associated with distinct $\xi_{j}$ characterizing internal, nonconservative response mechanisms. Like $\lambda$, $\xi$ is a macroscopic variable that acts in a similar manner across all energy levels. In Appendix A, we provide an illustrative example of its meaning for the system consisting of a volume of particles described in Fig.~1. The time dependence of all the aforementioned quantities arises from the temporal evolution of the internal parameter $\xi=\xi (t)$, which evolves dynamically according to its own kinetics, at constant values of $\beta$ and $\lambda$. At this stage, a few remarks are in order. First, Eqs.~(\ref{eq:Pi0}) and~(\ref{eq:Z}) should be interpreted as defining occupation probabilities and a partition function per particle for a system composed of $n$ particles. In the following and without loss of generality, we restrict the discussion to a single particle that can occupy one energy level at a time, each level being associated with an occupation probability. The remaining particles are treated as part of the thermal bath. In other words, we focus on the state of label $i$ of a single system within the ensemble (coarse-grained description), without resolving the underlying microscopic substructure (fine-grained description) \cite{TOLMAN}. This single-particle framework provides a simplified setting from which the extension to a system of $n$ particles (distinguishable or indistinguishable, e.g., fermions or bosons) can be carried out straightforwardly using standard methods of statistical mechanics \cite{TOLMAN}. For example, for distinguishable particles, the total partition function is given by the product of the single-particle partition functions over all $n$ particles (Maxwell–Boltzmann statistics) \cite{TOLMAN}. Each energy level corresponds to the energy of one accessible microstate in the extended canonical ensemble. Second, the extended canonical distribution considered here is interpreted as the nonequilibrium distribution involved during a stochastic experiment such as a work protocol. It remains valid for a small system both near equilibrium and far from equilibrium during a transformation. It provides a convenient means to account for time evolution occurring at constant values of other thermodynamic variables, through the time dependence of the transient energy levels. In this context, these energy levels should be interpreted as the energies of the accessible microstates, which become transient quantities with a finite lifetime when the system is out of equilibrium. Hereafter, we refer to these nonequilibrium energies of the accessible microstates as \textit{transient energy levels}. In the present framework, the departure from equilibrium is driven by the time dependence of an external control parameter associated with work exchange. Consequently, the origin of irreversibility is naturally attributed to the dynamics of these transient energy levels. This interpretation constitutes the foundation of the present theoretical framework.

\subsection{Identification of thermodynamic quantities in the extended nonequilibrium Gibbs canonical distribution}
We adopt a method originally introduced by Gibbs, which consists of perturbing the probability distribution function to identify the physically relevant terms \cite{GIBBS}. From Eqs.~(\ref{eq:Pi0}) and~(\ref{eq:Z}), the total differential of the state function $\Psi$ is
\begin{equation}
\label{eq:dPsi}
  d\Psi= \delta E-\frac{\delta S}{\beta k_{B}}+\frac{S}{\beta^2 k_{B}}d\beta.
\end{equation}
At constant temperature (i.e., fixed $\beta $), the infinitesimal variation of $\Psi$ arises from two sources: the variation of the energy and the entropy. The function $\Psi$ is a state function since it results from a sum over all accessible states. Its total differential includes contributions from variations in all independant variables $\beta$, $\lambda$, and $\xi$ but is itself a unique, well-defined quantity, independent of the index $i$. In contrast, the random variables $E$ and $S$ are not state functions. Consequently, we denote their infinitesimal variations with $\delta$, reflecting that they are path dependent at the microscopic level. Since the energy $E$ is now considered function of the external control variable $\lambda$ and the internal nonequilibrium variable $\xi$, its total variation under a change in these parameters can be expressed as
\begin{equation}
\label{eq:dEi}
  \delta E=\left( \frac{\delta E}{\partial \lambda} \right)_{\xi}d\lambda+ \left(  \frac{\delta E}{\partial \xi}  \right)_{\lambda}d\xi.
\end{equation}
The energy $E$ does not depend on $\beta$. However, the entropy $S$ (natural logarithm of the probability) depend on $\beta$ . Its infinitesimal variation can thus be written as
\begin{equation}
\label{eq:dSi2}
  \delta S=\left( \frac{\delta S}{\partial \beta} \right)_{\lambda, \xi}d\beta+\left( \frac{\delta S}{\partial \lambda} \right)_{\beta, \xi}d\lambda+ \left(  \frac{\delta S}{\partial \xi}  \right)_{\beta, \lambda}d\xi.
\end{equation}
Expanding the total differential of $\Psi$ as a sum of its partial derivatives with respect to the three state variables and identifying the partial derivatives, from Eq.~(\ref{eq:dPsi}) we write
\begin{subequations}
\begin{equation}
  \label{eq:beta}
  \left( \partial \Psi/\partial \beta \right)_{\lambda,\xi}=\frac{S}{\beta^2k_{B}}-\frac{\left( \delta S/\partial \beta \right)_{\lambda,\xi}}{\beta k_{B}} ,
\end{equation}
\begin{equation}
  \label{eq:lambda}
 \left( \partial \Psi/\partial \lambda \right)_{\beta,\xi}=\left( \delta E/\partial \lambda \right)_{\xi}-\frac{\left( \delta S/\partial \lambda \right)_{\beta,\xi}}{\beta k_{B}} ,
\end{equation}
\begin{equation}
  \label{eq:ksi}
   \left( \partial \Psi/\partial \xi \right)_{\beta,\lambda}=\left( \delta E/\partial \xi \right)_{\lambda}-\frac{\left( \delta S/\partial \xi \right)_{\beta,\lambda}}{\beta k_{B}} .
\end{equation}
\end{subequations}
Since work involves only variations in $\lambda$, from the second relation, Eq.~(\ref{eq:lambda}), the infinitesimal work exchanged between the system and the external bodies is defined as
\begin{equation}
\label{eq:dWi1}
  \delta W=\left( \delta E/\partial \lambda \right)_{\xi}d\lambda .
\end{equation}
For a single realization of an experiment, the infinitesimal work $\delta W$ measured on the small system can take on many possible values $\delta W_{i}$, depending on which state $i$ is realized at that moment. This definition corresponds to the notion of work in stochastic thermodynamics, where work is treated as a random variable. This is fully consistent with the standard framework of stochastic thermodynamics, provided the Hamiltonian of the system corresponds to the energy of the system, as in the canonical ensemble \cite{JAR1} 
\begin{equation}
\label{eq:dWi2}
  \delta W=\left(\partial \mathcal{H}/\partial \lambda \right)d\lambda .
\end{equation}
However, in our framework, changes in the energy are evaluated at constant values of the internal parameter $\xi$. As a result, the total work performed on the system during the transformation from an initial equilibrium state $\{A\}$ to a final state $\{B\}$ is given by the integral between $\lambda_{A}$ and $\lambda_{B}$ of the infinitesimal work contributions. It is important to emphasize that, in this approach, the total work is not equal to the difference of the Hamiltonian between states $\{A\}$ and $\{B\}$ because $\mathcal{H}$ (represented here by $E$) is not a state function. Here, $\mathcal{H}$ denotes the random variable energy of the system submitted to fluctuations, whereas only $\overline {\mathcal{H}}$ (or equivalently $\overline {E}=U$) corresponds to the internal energy of the system, which is a state function. The energy also depends on the internal parameter which evolves over time during the transformation. Thus, reaching the final equilibrium state $\{B\}$ requires considering not only the evolution $\lambda (t)$ but also the evolution in $\xi(t)$. From the third relation above, we now introduce a new quantity, defined as the random affinity, denoted by
\begin{equation}
  \label{eq:A1}
   A=-\left( \delta E/\partial \xi \right)_{\lambda} .
\end{equation}
For all microstates, $A_{i}$ reflects the internal forces associated with each energy level, arising from the nonequilibrium dynamics governed by $\xi(t)$. We can now express the random uncompensated heat of Clausius as
\begin{equation}
\label{eq:dQ'i1}
  \delta Q'=Ad\xi=-\left( \delta E \right)_{\lambda}  .
\end{equation}
This expression represents the microscopic analog of the De Donder formula for the uncompensated heat of Clausius in macroscopic nonequilibrium thermodynamics [see Eq.~(\ref{eq:diS1}] where $ \overline{\delta Q'} =\overline {A}d\xi$). Within this approach, the entropy production acquires a clear and physically grounded meaning at the microscopic level
\begin{equation}
\label{eq:diS3}
  \delta_{i}S=\beta k_{B}Ad\xi=-\beta k_{B}\left( \delta E \right)_{\lambda} .
\end{equation}
At the microscopic scale, entropy production is directly associated with variations in the random energy and, consequently, in the transient energy levels as the internal variable $\xi$ evolves. For a given state, the corresponding infinitesimal entropy production may take either positive or negative values. However, in accordance with the second law of thermodynamics, its statistical average over all possible realizations is always positive. This is a formulation in which both an affinity and an uncompensated heat of Clausius are defined at the microscopic level as random quantities. Finally, the mean work supplied to the system and the mean uncompensated heat produced during an infinitesimal variation process are expressed, respectively, as
\begin{subequations}
\begin{align}
  \label{eq:dWmean}
  \overline{\delta W} & =\sum_{i=1}^{N}P_{i}\delta W_{i}= \sum_{i=1}^{N}P_{i}\left( \delta E_{i}/\partial \lambda \right)_{\xi}d\lambda \nonumber \nonumber\\ 
  &=-\frac{1}{\beta}\left( \partial \ln Z/\partial \lambda \right)_{\beta,\xi}d\lambda=\left( \partial F/\partial \lambda \right)_{\beta,\xi}d\lambda ,
\end{align}
\begin{align}
  \label{eq:dQ'mean}
  \overline{\delta Q'} & =\sum_{i=1}^{N}P_{i}\delta Q'_{i}=\sum_{i=1}^{N}P_{i}A_{i}d\xi= -\sum_{i=1}^{N}P_{i}\left( \delta E_{i}/\partial \xi \right)_{\lambda}d\xi\nonumber \\ 
  &=\frac{1}{\beta}\left( \partial \ln Z/\partial \xi \right)_{\beta,\lambda}d\xi=- \left( \partial F/\partial \xi \right)_{\beta,\lambda}d\xi.
\end{align}
\end{subequations}
These expressions correspond precisely to the definitions of work and uncompensated heat in macroscopic nonequilibrium thermodynamics. We recover the expressions given by Eqs.~(\ref{eq:dW1}) and~(\ref{eq:f}) (evaluated here at constant $\xi$) as well as Eq.~(\ref{eq:diS1}), which defines the entropy production in terms of the thermodynamic affinity $\overline{A}=- \left( \partial F/\partial \xi \right)_{\beta,\lambda}$. In Appendix A, where this framework is applied to the system in Fig.~1, the microscopic affinity is defined as the pressure difference between the internal and external pressures with respect to the vessel, while the internal variable $\xi$ is the volume of the gas particles. With these microscopic definitions in place and noting that the average variation of the random entropy vanishes due to the normalization constraint on the probabilities
\begin{equation}
\label{eq:dSiCancel}
  \overline{\delta S}=-k_{B}\sum_{i=1}^{N}\delta P_{i}=0 ,
\end{equation}
then averaging Eq.~(\ref{eq:dPsi}) over the nonequilibrium canonical ensemble yields
\begin{align}
\label{eq:dPsi2}
  d\Psi = dF & =\overline{\delta E}+\frac{\overline{S}}{\beta^2 k_{B}}d\beta \nonumber \\
   & =\overline{\left( \delta E/\partial \lambda \right)_{\xi}}d\lambda                       + \overline{\left( \delta E/\partial \xi \right)_{\lambda}}d\xi +\frac{\overline{S}}{\beta^2 k_{B}}d\beta \nonumber \\
  & = \overline{\delta W} - \overline{\delta Q'} + \frac{\overline{S}}{\beta^2 k_{B}}d\beta .
\end{align}
This expression is formally identical to Eq.~(\ref{eq:dQ'2}) from macroscopic nonequilibrium thermodynamics, provided that the system temperature is identified with $T=1/k_{B}\beta$, where $\beta$ is, as previously recalled, the inverse of the modulus in the canonical Gibbs distribution \cite{GIBBS}. $\overline{S}$ is the macroscopic entropy defined in Sec. II. To summarize, within our framework, both the random work exchanged between the system and its surroundings and the random uncompensated heat of Clausius acquire clear meanings
\begin{subequations}
\begin{equation}
  \label{eq:dWi3}
  \delta W=\left( \delta E \right)_{\xi} ,
\end{equation}
\begin{equation}
  \label{eq:dQ'i3}
  \delta Q'=-\left( \delta E \right)_{\lambda}.
\end{equation}
\end{subequations}
For isothermal transformations ($\beta=\mathrm{const}$), the variation of the random energy coincides with the variation of the free energy, up to contributions arising from changes in the probability distribution (i.e., in the statistical entropy)
\begin{equation}
\label{eq:dPsi3}
 \delta E=\delta W-\delta Q'=d\Psi+\frac{\delta S}{\beta k_{B}}  .
\end{equation}
This relation forms the basis for the subsequent derivation of nonequilibrium relations.

\section{Nonequilibrium relations}
In this section, we first recover the nonequilibrium work relation by means of the fluctuation relation for the entropy production. We then derive a new nonequilibrium relation for the random heat associated with the random work performed during the work protocol. 

\subsection{Nonequilibrium work relation or Jarzynski equality}
Let us rewrite Eq.~(\ref{eq:dPsi}), assuming $\beta =$ constant, as follows:
\begin{equation}
\label{eq:dPsi4}
 \beta d\Psi-\beta \delta E=\delta \ln P  .
\end{equation}
The nonequilibrium work protocol consists of varying the external control parameter $\lambda$ at a constant rate $d\lambda/dt$ over a finite time interval $\tau$ from an inital value $\lambda_A$ to a final value $\lambda_B$  \cite{JAR2}. During this process, the system is driven from an initial equilibrium state $\{A\}(\beta, \lambda_A)$ into a nonequilibrium transient state, and subsequently, after relaxation, it reaches a final equilibrium state $\{B\}(\beta, \lambda_B)$. At the initial state $\{A\}$, due to equilibrium, the parameter $\lambda_A$ depends solely on the thermodynamic equilibrium variable $\xi_A=\xi_A^{\mathrm{eq}}$, i.e., $\lambda_A = f(\xi_A)$. The same holds for the final state $\{B\}$. Since we consider only isothermal transformations we omit explicit dependence on $\beta$. However, during the transformation between $\{A\}$ and $\{B\}$, the system is generally out of equilibrium, and the external parameter $\lambda$ becomes a function not only of $\xi$, but also of its conjugate affinity $A$; that is, $\lambda = f(\xi, A)$ (see Appendix C). Since $\{A\}$ and $\{B\}$ are equilibrium states, the integral of $d\Psi$ from the state function $\Psi$ (which we shall denote by $F$ from now on, as it corresponds to the Helmholtz free energy) is path-independent. It depends solely on the intrinsic properties of the system in $\{A\}$ and in $\{B\}$. Therefore, without loss of generality, we can consider a specific nonequilibrium path, among others, connecting the two equilibrium states. The essential point is that, since the observable is obtained from an average over all realizations of the relevant random variables, the particular nonequilibrium trajectory is irrelevant provided the initial and final equilibrium states are unchanged and the rate of variation of $\lambda$ is the same for all realizations. This illustrates the power of nonequilibrium relations, such as the Jarzynski equality, whose scope is remarkably broad since they apply to any type of nonequilibrium transformation, whether the system remains close to or far from equilibrium during the process. We thus consider a two-step protocol such as depicted in Fig.~2. This two-step process is commonly used in stochastic thermodynamics (see, for example, Appendix A, where the model system in Fig.~1 is subjected to the two-step work protocol, and references therein).
\begin{figure}[!htbp]
  \centering
  \includegraphics[width=8.6 cm]{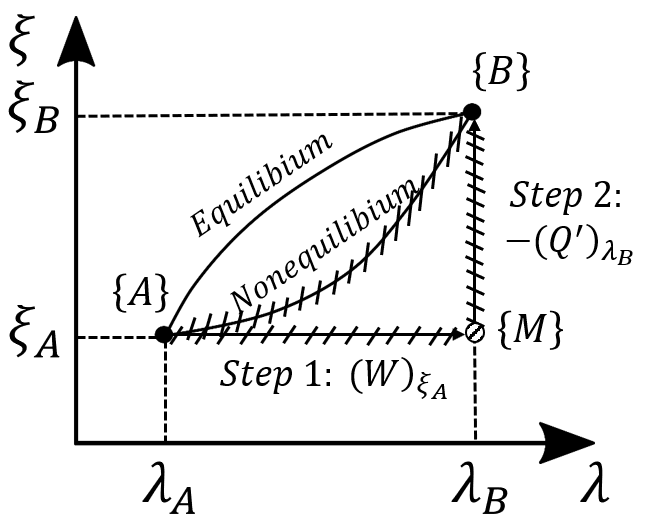}
  \caption{Transformations between equilibrium states $\{A\}$ and $\{B\}$ in the $(\lambda,\xi)$ plane. The unique reversible (equilibrium) path is shown as a solid black line. Nonequilibrium trajectories are indicated by solid black lines with hatching, emphasizing that these trajectories are not uniquely determined. One representative nonequilibrium trajectory consists of two successive steps: first, step 1 at constant $\xi$, followed by step 2 at constant $\lambda$.}
  \label{fig:Fig2}
\end{figure}
In step 1, $\lambda$ is varied at constant rate from $\lambda_{A}$ to $\lambda_{B}$ while keeping $\xi$ fixed at $\xi_{A}$. Since $\xi$ remains constant, only work is performed during this stage. This is like an equilibrium path but at constant $\xi$. However, at the end of step 1, the system is no longer in equilibrium; its state is characterized by $\lambda_{B}$ and $\xi_{A}$. This intermediate, nonequilibrium state is denoted by $\{M\}$ in Fig.~2. This situation is analogous to a sudden quench, where the system becomes trapped in a frozen-in, nonequilibrium state. Although no uncompensated heat is produced during this step, the affinity is non-zero because $\lambda = f(\xi, A)$ (see Appendix C). In other words, at time $\tau$, there are thermodynamic forces acting on the system, but the corresponding fluxes vanish on average due to the frozen condition. This is consistent with the assumption that $\xi$ is held fixed. In step 2, the variable $\xi$ is allowed to evolve from $\xi_{A}$ to $\xi_{B}$ at fixed $\lambda = \lambda_{B}$. This step is necessary for the system to relax toward the final equilibrium state $\{B\}$, where $\lambda_{B} = f(\xi_{B})$ and the affinity vanishes (see Appendix C). The relaxation occurs over a finite time interval, governed by the microscopic relaxation times that reflects the mechanical disequilibrium at the microscopic scale. Notably, step 1 is only possible under the condition that the typical macroscopic relaxation time of the nonequilibrium processes is much greater than the switching time $\tau$. As shown in Fig.~2, a unique equilibrium path (solid black line) connects the two equilibrium states. On this path, the system satisfies the relation $\xi =\xi^{\mathrm{eq}}=g(\lambda)$, reflecting the fact that $\xi$ and $\lambda$ are not independent in equilibrium. This relation corresponds to the equation of state of the system at constant temperature. In contrast with the two-step process (or other nonequilibrium paths such as that shown in Fig.~2), the existence of this equilibrium pathway requires that the switching time $\tau$ be much longer than the characteristic macroscopic relaxation time of the system. During step 2, uncompensated heat (and corresponding entropy production) is generated, but no work is done since $\lambda$ remains constant. \\ 
In Fig.~3, we illustrate qualitatively this two-step process using the transient energy-level framework based on the extended canonical distribution, applied to a simplified three-level system for clarity. 
\begin{figure}[!htbp]
  \centering
  \includegraphics[width=8.6 cm]{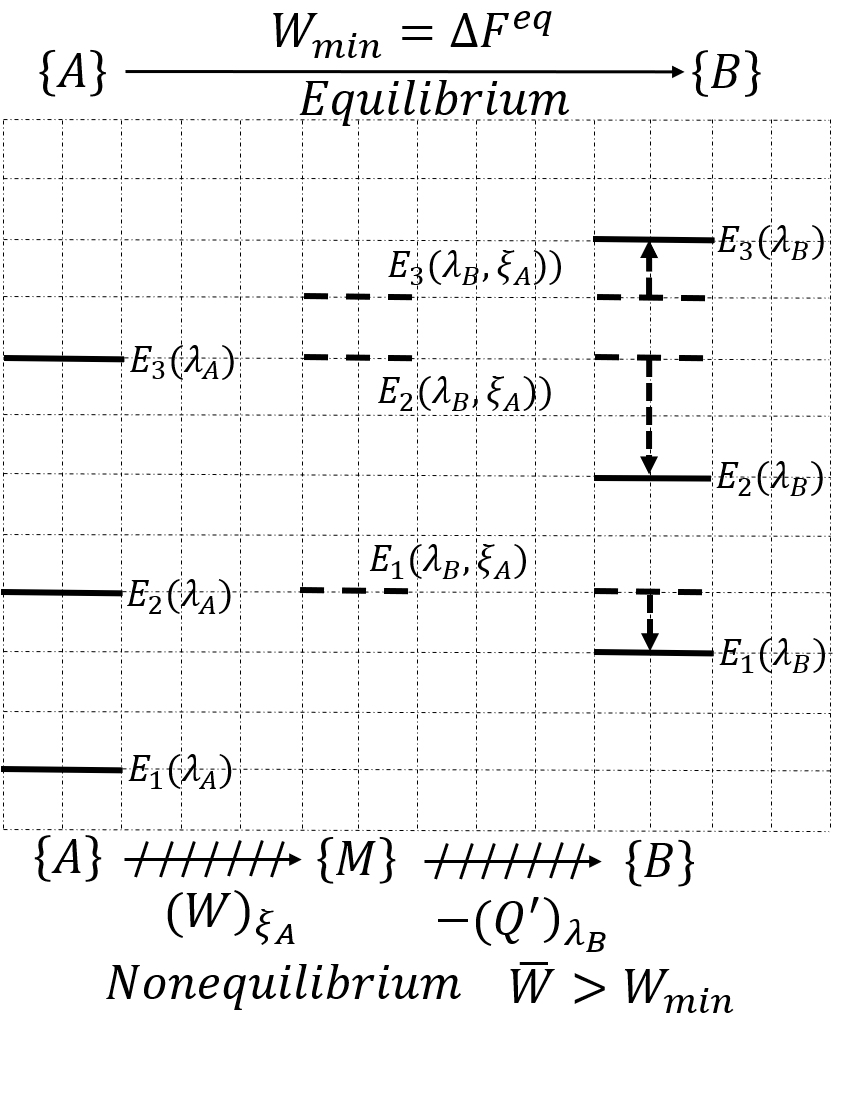}
  \caption{Evolution along the two-step path, illustrated qualitatively in terms of transient energy levels for a simplified three-level system. The equilibrium energy levels of states $\{A\}$ and $\{B\}$ are shown as solid black lines. The transient energy levels of the intermediate nonequilibrium state $\{M\}$ are depicted by dashed black lines. Relaxation of these transient levels toward the equilibrium levels of state $\{B\}$ is indicated by a dashed black arrow.}
  \label{fig:Fig3}
\end{figure}
For an equilibrium transformation under external work, the three energy levels of state $\{A\}$ evolve over time into the corresponding levels of state $\{B\}$, with the minimum possible work performed on the system. In this case, represented by the equilibrium line in Fig.~2, the average work is equal to the free-energy difference between equilibrium states $\{A\}$ and $\{B\}$. For the two-step nonequilibrium trajectory, this is no longer the case. During step 1 the work fluctuates according to $\left ( W_{i}\right )_{\xi_{A}}=E_{i}(\lambda_{B},\xi_{A})-E_{i}(\lambda_{A})$ from an initial microstate $i$ randomly sampled in state $\{A\}$. In this case, the average work exceeds the equilibrium free-energy difference, as the transient energy levels are elevated with respect to their equilibrium values in state $\{B\}$, except for level 3 in our example. Although unlikely, this event has a significant impact on nonequilibrium relations. In step 2, uncompensated heat is generated and the transient energy levels relax toward their equilibrium values with their own kinetics (recall that transient energy levels refers to the time-dependent energies of the accessible microstates in the nonequilibrium state). In this qualitative example, however, level 3 exhibits an apparent negative uncompensated heat, relaxing toward a higher energy value. Such events occur with low probability, but the average uncompensated heat remains positive, as required by the second law. This is directly linked to the fact that $(\Delta E_{3})_{\xi_{A}}=E_{3}(\lambda_{B},\xi_{A})-E_{3}(\lambda_{A})$ is smaller than $\Delta F^{\mathrm{eq}}$ for this level. Although rare, such fluctuations play a dominant role in nonequilibrium relations, as they strongly affect the exponential averaging. Events of this type lie in the distribution tail and are thus extremely rare, requiring a sufficiently large number of realizations for their occurrence to be observed. This two-step protocol has the advantage of clearly separating the two random variables contributing to the energy change $\delta E$ along an entire trajectory
\begin{equation}
\label{eq:DeltaE}
 \int \delta E= \left( W \right)_{\xi}- \left( Q' \right)_{\lambda}.
\end{equation}
Considering step 1, the total work performed on the system as the control parameter $\lambda$ varies from $\lambda_{A}$ to $\lambda_{B}$ is given by
\begin{equation}
\label{eq:dW4}
  W= \int_{\lambda_{A}}^{\lambda_{B}}\delta W=\int_{\lambda_{A}}^{\lambda_{B}}\left( \delta E \right)_{\xi_{A}}.
\end{equation}
Since $\xi$ is constant, we can integrate Eq.~(\ref{eq:dPsi4}) along the horizontal line in Fig.~2
\begin{align}
\label{eq:dPsi5}
\beta (\Delta F)_{\xi_{A}}
 - \beta \left( W \right)_{\xi_{A}}
 &= \int_{\lambda_{A}}^{\lambda_{B}}
    \left( \delta \ln P \right)_{\xi_{A}} \nonumber \\
&= \ln \left(
    \frac{P(\lambda_{B})_{\xi_{A}}}{P(A)}
   \right).
\end{align}
Let $(\Delta F)_{\xi_{A}}$ denote the Helmholtz free-energy change associated with the system on step 1. The probability of the initial equilibrium state $\{A\}$ is denoted by $P(A)=P[\lambda_{A}(\xi_{A})]$, while $P(\lambda_{B})_{\xi_{A}}$ represents the probability reached at the end of step 1 under the condition that $\xi = \xi_{A}$. It is important to note that, in step 1, we can integrate $\delta \ln P$, even though it does not correspond to the partial derivative of a total differential, since only one variable is varied during this process. In the general case, such an integration would be path dependent. In other words, at the end of step 1 only a part of the total phase space has been explored by the system in the switching time $\tau$. Taking the exponential of both sides of the resulting expression and then averaging over many realizations of step 1, each initiated from the equilibrium probability $P_i(A)$ associated with the microscopic state $i$ in the initial macroscopic equilibrium state $\{A\}$, we obtain
\begin{equation}
\label{eq:ExpdPsi6}
 e^{\beta (\Delta F)_{\xi_{A}}}\sum_{i=1,\xi_{A}}^{N} P_{i}(A)e^{-\beta \left ( W_{i}\right )_{\xi_{A}}} = \sum_{i=1,\xi_{A}}^{N}P_{i}(\lambda_{B})_{\xi_{A}}=1 .
\end{equation}
In this case, the averaging procedure is restricted to a subensemble of the phase space in which $\xi$ remains constant and equal to $\xi_{A}$. On this subensemble we have
\begin{equation}
\label{eq:ExpdPsi62}
 \sum_{i=1,\xi_{A}}^{N} P_{i}(A)e^{-\beta \left ( W_{i}\right )_{\xi_{A}}} = e^{-\beta (\Delta F)_{\xi_{A}}} .
\end{equation}
This is the nonequilibrium work relation limited to the accessible states with $\xi=\xi_{A}$. The free-energy difference $(\Delta F)_{\xi_{A}}$ refers to the change obtained at the end of step 1, and the system is still in a nonequilibrium state (state $\{M\}$ in Fig.~2 and~3). During step 2, the total uncompensated heat generated in the system as $\xi$ evolves from $\xi_A$ to $\xi_B$ is given by
\begin{equation}
\label{eq:dQ'4}
  Q'= \int_{\xi_{A}}^{\xi_{B}}\delta Q'=-\int_{\xi_{A}}^{\xi_{B}}\left( \delta E \right)_{\lambda_{B}}=\int_{\xi_{B}}^{\xi_{A}}\left( \delta E \right)_{\lambda_{B}}.
\end{equation}
This step involves only entropy production, with no work being performed. It is appropriate, in this case, to consider the time-reversed process, which begins at the final equilibrium state $\{B\}$ and evolves backward in time toward the intermediate nonequilibrium state $\{M\}$. Since this reversed process starts from an equilibrium state, we can apply Crooks formula for entropy production \cite{CROOKS1} or the fluctuation theorem \cite{EVANS1,EVANS2}
\begin{equation}
\label{eq:Crooks1}
 \frac{P_{F}(+\omega)}{P_{R}(-\omega)}=e^{+\omega} .
\end{equation}
Here $P_F(+\omega)$ denotes the probability of observing an entropy production $+\omega$ along a forward trajectory, while $P_R(-\omega)$ represents the probability of observing the negative entropy production $-\omega$ along the corresponding time-reversed trajectory, where all momenta are reversed in phase space. In the context of our time-reversed step 2, the forward process corresponds to the evolution from the equilibrium state $B$, where $\xi = \xi_B$ and $\lambda = \lambda_B$, to the nonequilibrium state $\{M\}$, characterized by $\xi = \xi_A$ and $\lambda = \lambda_B$. In this case, the Crooks fluctuation theorem takes the form
\begin{equation}
\label{eq:Crooks2}
 \frac{P(B)}{P(\xi_{A})_{\lambda_{B}}}=e^{ \beta \left ( Q'\right )_{\lambda_{B}}} .
\end{equation}
The probability $P(\xi_{A})_{\lambda_{B}}$ is the probability reached at the end of the reverse step 2 under the condition $\lambda=\lambda_{B}$. For this reverse step 2, after integration of Eq.~(\ref{eq:dPsi4}) we have
\begin{align}
\label{eq:dPsi7}
-\beta (\Delta F)_{\lambda_{B}}
 - \beta \left( Q' \right)_{\lambda_{B}}
 &= \int_{\xi_{B}}^{\xi_{A}}
    \left( \delta \ln P \right)_{\lambda_{B}} \nonumber \\
&= \ln \left(
    \frac{P(\xi_{A})_{\lambda_{B}}}{P(B)}
   \right).
\end{align}
The integration has been performed along the vertical line in Fig.~2. Another subvolume of the phase space has been explored by the system under the constraint that $\lambda = \lambda_{B}$. Also, remember that $ \left( \delta E \right)_{\lambda_{B}}=-\left ( \delta Q'\right )_{\lambda_{B}}$. Taking the exponential of this expression and performing a statistical average over many realizations of reverse step 2, each initiated from the equilibrium probability $P_i(B)$ associated with the microscopic state $i$ in the macroscopic equilibrium state $\{B\}$, we obtain
\begin{align}
\label{eq:dPsi8}
e^{-\beta (\Delta F)_{\lambda_{B}}}\sum_{i=1,\lambda_{B}}^{N}
   P_{i}(B) \, e^{-\beta \left( Q'_{i} \right)_{\lambda_{B}}}
&= \sum_{i=1,\lambda_{B}}^{N}
   P_{i}(\xi_{A})_{\lambda_{B}} \nonumber \\
&= 1.
\end{align}
As in step 1, we obtain a nonequilibrium relation for uncompensated heat this time, limited to a part of the total phase space for which $\lambda = \lambda_{B}$
\begin{equation}
\label{eq:dQ'i14}
 \sum_{i=1,\lambda_{B}}^{N}P_{i}(B) e^{-\beta \left ( Q'_{i}\right )_{\lambda_{B}}}=e^{\beta (\Delta F)_{\lambda_{B}}}
 .
\end{equation}
However, from Eq.(\ref{eq:Crooks2}), we notice that the left term in the previous equation is equal to 1 on this limited volume of the phase space. The free-energy difference $(\Delta F)_{\lambda_{B}}$ is therefore exactly zero during step 2 (or in the reverse step 2 since $F$ is a state function). When considering statistics over the full ensemble of combined step 1 and 2 processes, the total free-energy difference is given by $(\Delta F)_{\xi_{A}}+(\Delta F)_{\lambda_{B}}=(\Delta F)_{\xi_{A}}=\Delta F^{\mathrm{eq}}$, as the initial and final states are both at equilibrium and $F$ is a state function. Consequently, over a series of step 1 + step 2,  Eq.~(\ref{eq:ExpdPsi6}) becomes for the complete process
\begin{align}
\label{eq:jarzynski}
 \sum_{i=1,\lambda_{B}}^{N}P_{i}(\xi_{A})_{\lambda_{B}}& \sum_{i=1,\xi_{A}}^{N} P_{i}(A) e^{-\beta \left ( W_{i}\right )_{\xi_{A}}}\nonumber \\ &= \overline{e^{-\beta W}} \nonumber \\ &= \sum_{i=1,\lambda_{B}}^{N}P_{i}(\xi_{A})_{\lambda_{B}} e^{-\beta (\Delta F)_{\xi_{A}}} \nonumber \\ &=e^{-\beta \Delta F^{\mathrm{eq}}}.
\end{align}
This is the nonequilibrium work relation \cite{JAR2}. The overline denotes an average taken over the full phase space explored by the system or, equivalently, over all accessible microscopic states of the system. This confirms, as noticed by Jarzynski, that the equilibrium stage (step 2) is somewhat superfluous for establishing this equality \cite{JAR1}. It is worth noting that no additional contribution to the free energy between the two equilibrium states arises during the relaxation process. During step 2, it is as though all the available work has been converted into uncompensated heat [i.e., $(\Delta F)_{\lambda_{B}} = 0$], as no further work can be performed once $\lambda$ is held fixed. This follows directly from the fundamental relation~(\ref{eq:Crooks1}) governing entropy production.  However, for the statistics to be carried out over the complete phase space of the system, it is necessary that step 2 be fulfilled. It is crucial that the system reaches equilibrium in $\{B\}$ for $(\Delta F)_{\xi_{A}} = \Delta F^{\mathrm{eq}}$ to hold.  Let us note that from Eq.~(\ref{eq:dPsi8}), a nonequilibrium relation for the uncompensated heat is obtained, namely \cite{CROOKS2, EVANS3, CROOKS1}
\begin{align}
\label{eq:dQ'5}
\sum_{i=1,\xi_{A}}^{N}P_{i}(\lambda_{B})_{\xi_{A}}& \sum_{i=1,\lambda_{B}}^{N}P_{i}(B) e^{-\beta \left ( Q'_{i}\right )_{\lambda_{B}}}\nonumber \\ &= \overline{e^{-\beta Q'}} \nonumber \\ &=1 .
\end{align}
The averaging has been performed from the equilibrium state $B$ with the probability $P_{i}(B)$ on the entire phase space of the system. Here, $\Delta_{i} S = Q'/T$ is the total entropy production on a stochastic process of step 1 + step 2. In the general case (unlike in our specific two-step transformation of step 1 followed by step 2) the parameters $\lambda$ and $\xi$ evolve simultaneously in disturbing the distribution of the energy levels at equilibrium. This is because $\lambda$ and $\xi$ are independent state variables, even though the transformation is induced by a variation in $\lambda$ from the oustside (see Appendix C). Nevertheless, both nonequilibrium relations remain valid since the process connects the same two equilibrium states. The nonequilibrium relations in Eqs.~(\ref{eq:jarzynski}) and~(\ref{eq:dQ'5}) are therefore intrinsically linked, and the former cannot be derived without invoking the latter. This is a prerequisite for the system to reach the equilibrium state $\{B\}$ starting from the equilibrium state $\{A\}$ after having evolved over the full phase space according to the equations of motion. The fluctuations of work and uncompensated heat are thus completely entangled. Their connection can be further highlighted as follows. From the equalitiesin Eqs.~(\ref{eq:lambda}) and~(\ref{eq:ksi}), we identify the fluctuations of work and uncompensated heat along a path
\begin{subequations}
\begin{equation}
  \label{eq:FluctuW}
 \Delta (W)= W-\overline{W}=\frac{\int {\left( \delta S \right)_{\xi}} }{\beta k_{B}},
\end{equation}
\begin{equation}
  \label{eq:FluctuQ'}
   -\Delta (Q')= -Q'+\overline{Q'}=\frac{\int {\left( \delta S \right)_{\lambda}}}{\beta k_{B}}.
\end{equation}
\end{subequations} 
We adopt the standard definition of a fluctuation as the deviation of a random variable from its mean value \cite{GIBBS}. Accordingly, the two fluctuations are connected through the extended canonical distribution [Eq.~(\ref{eq:Pi0})] with $S=-k_{B}\ln P$. For a system with discrete energy levels, this means that once the energy spectrum is specified together with its dependence on the control parameter $\lambda$ and once the relaxation dynamics toward equilibrium are characterized for the variable $\xi$, the extended canonical probability distribution can be driven out of equilibrium by the applied work protocol. This, in turn, governs the evolution of the occupation probabilities (or statistical entropies), which encode the fluctuations of both the random work and the random uncompensated heat in the system. Taking the average of the two fluctuations introduced above reveals that their mean values vanish simultaneously with $\overline{\int{\delta S}}=0$ whereas the mean values of work and uncompensated heat obey the second law of thermodynamics, following $\overline{W} - \overline{Q'} = \Delta F^{\mathrm{eq}}$. In conclusion, the extended canonical probability distribution fully determines the fluctuations of both work and uncompensated heat throughout the entire transformation from an initial equilibrium state $\{A\}$ to a final equilibrium state $\{B\}$, as prescribed by the work protocol. Figure~4 schematically illustrates the respective fluctuations of the random work and the random uncompensated heat of Clausius around their mean values.
\begin{figure}[!htbp]
  \centering
  \includegraphics[width=8.6 cm]{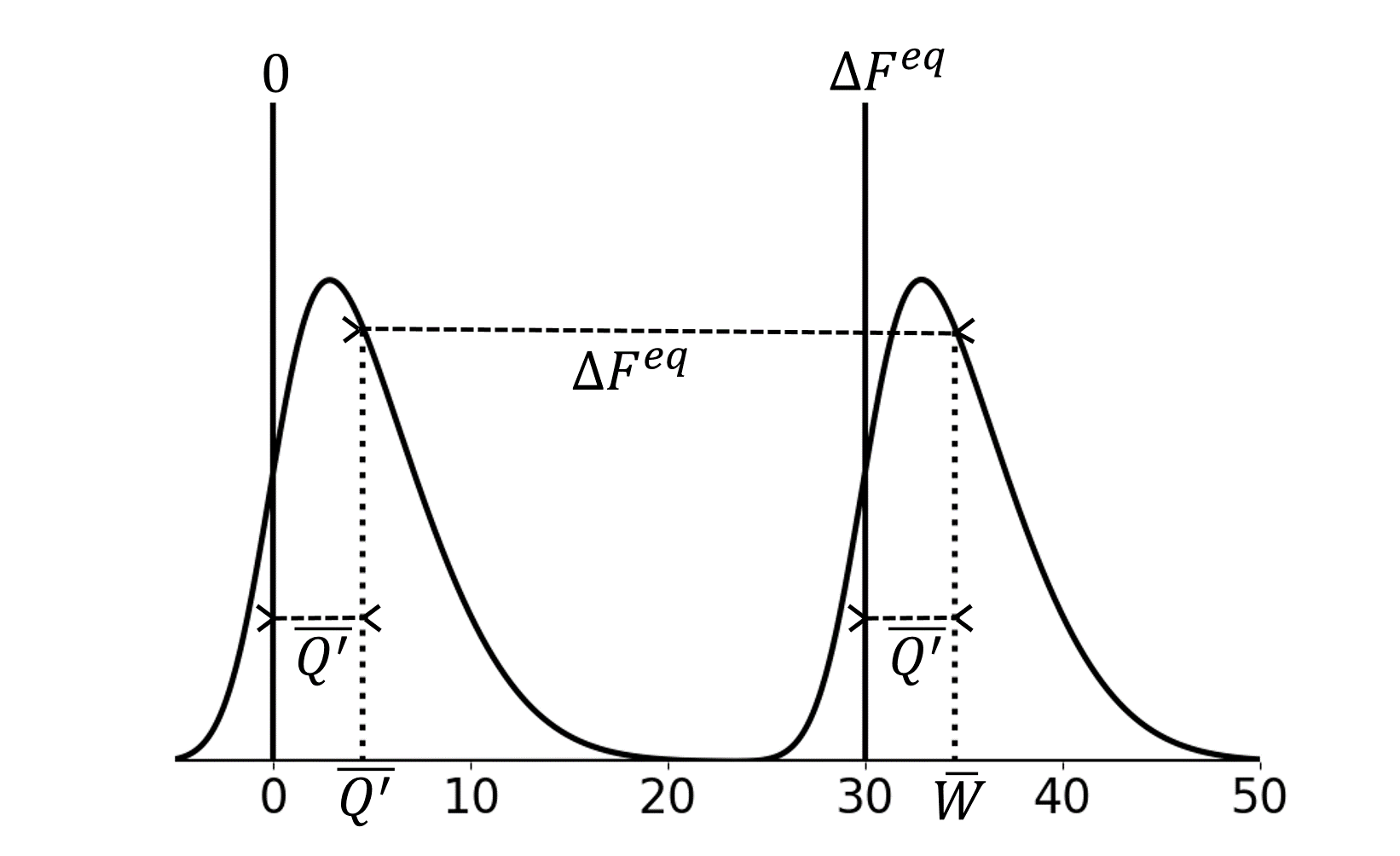}
  \caption{The random work and random uncompensated heat of Clausius are distributed around their mean values (dot lines) in the same graph. The abscisse represents energy in arbitrary units of Joule. See text for explanation. }
  \label{fig:Fig4}
\end{figure}
The mean value of the uncompensated heat is strictly positive, as required by the second law of thermodynamics. It is distributed around this mean, yet it exhibits a tail extending into negative values, reflecting fluctuations within the ensemble. Similarly, the mean work exceeds the equilibrium Helmholtz free-energy difference $\Delta F^{\mathrm{eq}}$, in accordance with the second law. In this case fluctuations allow for a tail of the distribution extending into values smaller than $\Delta F^{\mathrm{eq}}$ across the ensemble. The difference between the mean work and $\Delta F^{\mathrm{eq}}$ exactly matches the difference between the mean uncompensated heat and its equilibrium value, which is zero. To clarify the connection between the two curves, it is useful to consider a transformation in which the change in internal energy exactly compensates the change in entropy, $\Delta U^{\mathrm{eq}}=T\Delta S^{\mathrm{eq}}$, that is, a transformation without variation of free energy. In this case, the two curves shown in Fig.~4 coincide, with $\overline{ W}=\overline{ Q'}$.

\subsection{Nonequilibrium heat relation }  
In all that precedes, the quantity of heat has not yet been defined at the microscopic level. However, the amount of heat has been introduced macroscopically in the section on macroscopic nonequilibrium thermodynamics. Indeed, heat and work involved in a process are related through the first law of thermodynamics [Eq.~(\ref{eq:FirstLaw1})]. We may assume that the existence of a nonequilibrium relation for work implies the existence of a corresponding nonequilibrium relation for the amount of heat transferred to the heat bath. At this point, it is important to emphasize that the uncompensated heat of Clausius (interpreted in our approach as being associated with the relaxation of transient energy levels in a small system) is an internal property of the system. In other words, uncompensated heat is an amount of entropy produced within the boundaries that define the system, whereas heat, in contrast, is a transfer of entropy across those boundaries between the system and its surroundings. On average, the uncompensated heat corresponds to the potential amount of entropy that did not have time to be compensated by an exchange with the heat bath during the work protocol (hence the term uncompensated heat, as introduced by Clausius). A practical way to highlight the distinction between heat and uncompensated heat of Clausius is to consider an adiabatic transformation (where no heat is exchanged with the surroundings) in which the uncompensated heat accounts for the entropy irreversibly produced within the system, without any compensation by the environment. In this case, the system temperature increases, and the transformation is not isothermal. We must now define what is meant by heat at the microscopic scale, under the assumption that, on average, this definition is consistent with the macroscopic one provided by thermodynamics. Nevertheless, at the microscopic scale, the notion of heat is not uniquely defined, and several distinct definitions coexist in the literature \cite{SEKI1}. Here, we analyze heat within the framework of the first law of thermodynamics [(Eq.~(\ref{eq:FirstLaw1})]. At the microscopic level, the infinitesimal amount of heat corresponds to the difference between the total differential of the state function $U = \overline{E}$ and the random work exchanged with the surroundings
\begin{equation}
\label{eq:dQi1}
\delta Q = dU - \delta W .
\end{equation}
Since $dU$ is a total differential, the heat exchange is also a random variable, as the work is itself a random variable for small systems. Using the fundamental relation in Eq.~(\ref{eq:dPsi3}), derived from the extended canonical distribution, the random work can be expressed in terms of the free energy, yielding
\begin{equation}
\label{eq:dQi2}
\delta Q = \frac{d\overline{S}}{\beta k_{B}} - \delta Q'-\frac{\delta S}{\beta k_{B}} .
\end{equation}
Equation~(\ref{eq:dQi1}) corresponds to the conservation of energy at the microscopic scale, whereas Eq.~(\ref{eq:dQi2}) expresses the second law at the microscopic scale. The entropy $\overline{S}$ is the average of the statistical entropy or random entropy $S$. Taking the statistical average of both relations, we recover the first law of thermodynamics [Eq.~(\ref{eq:FirstLaw1})] and the second law of thermodynamics [Eq.~(\ref{eq:heat2})], provided that $\overline{\delta S}=0$ as required by the normalization of the probability distribution. By rewriting the latter relation, we establish a connection between the random heat and the random uncompensated heat (entropy production) in terms of entropy fluctuations
\begin{equation}
\label{eq:dQi3}
\delta Q+ \delta Q' = \frac{d\overline{S}-\delta S}{\beta k_{B}}=\frac{\delta \chi}{\beta k_{B}} .
\end{equation}
Indeed, if we define the random variable $\chi = \overline{S} - S$ as the fluctuation of entropy, then $\delta \chi=d\overline{S} - \delta S$ simply represents the variation of the entropy fluctuation during an infinitesimal process. Since the random entropy is not a state function and therefore cannot be directly integrated along an entire nonequilibrium path, we again consider the two-step nonequilibrium path introduced in the previous section. We first consider step 1, as in the previous section. The integration over $\lambda$ in this step is carried out with $\xi = \xi_{A}$  where Eq.~(\ref{eq:dQi2}) becomes
\begin{equation}
\label{eq:dQi4}
\beta\left(Q \right)_{\xi_{A}}- \frac{\left(\Delta \overline{S} \right)_{\xi_{A}}}{k_{B}}= \ln \left(\frac{P(\lambda_{B})_{\xi_{A}}}{P(A)} \right) .
\end{equation}
Since $\xi$ is constant, there is no entropy production in this step. Taking the exponential of both sides of the resulting expression and performing a statistical average over many realizations of step 1, each initialized from the equilibrium distribution $P_{i}(A)$, we obtain
\begin{align}
\label{eq:dQi5}
 \sum_{i=1,\xi_{A}}^{N} P_{i}(A) e^{\beta\left(Q_{i} \right)_{\xi_{A}}} e^{-\frac{\left(\Delta \overline{S} \right)_{\xi_{A}}}{k_{B}}}&=\sum_{i=1,\xi_{A}}^{N}P_{i}(\lambda_{B})_{\xi_{A}}\nonumber \\ &=1.
\end{align}
Once again, the statistical averaging is performed over the restricted subphase space of step 1. We therefore obtain under this condition
\begin{equation}
\label{eq:dQi6}
 \sum_{i=1,\xi_{A}}^{N} P_{i}(A) e^{\beta\left(Q_{i} \right)_{\xi_{A}}}=e^{\frac{\left(\Delta \overline{S} \right)_{\xi_{A}}}{k_{B}}}.
\end{equation}
We next consider the reverse step 2, as in the previous section, where $\xi$ varies from $\xi_{B}$ to $\xi_{A}$ while $\lambda$ is kept fixed at $\lambda_{B}$ for the integration of Eq.~(\ref{eq:dQi2})
\begin{equation}
\label{eq:dQi7}
 \beta\left(Q \right)_{\lambda_{B}}- \frac{\left(\Delta \overline{S} \right)_{\lambda_{B}}}{k_{B}}= -\beta\left(Q' \right)_{\lambda_{B}}+\ln \left(\frac{P(B)}{P(\xi_{A})_{\lambda_{B}}} \right) .
\end{equation}
In this step, entropy is produced (we also recall that no work is performed). Taking the exponential of both sides of the resulting expression and performing a statistical average over many realizations of reverse step 2, each initialized from the equilibrium distribution $P_{i}(B)$, we obtain
\begin{align}
\label{eq:dQi8}
 \sum_{i=1,\lambda_{B}}^{N}P_{i}(\xi_{A})_{\lambda_{B}}& e^{\beta\left(Q_{i} \right)_{\lambda_{B}}} e^{-\frac{(\Delta \overline{S})_{\lambda_{B}}}{ k_{B}}}& \nonumber \\ =\sum_{i=1,\lambda_{B}}^{N} P_{i}(B)& e^{-\beta\left(Q'_{i} \right)_{\lambda_{B}}}.
\end{align}
Statistical averaging is once again performed over the restricted subphase space of step 2. In this subvolume of phase space, we apply Eq.~(\ref{eq:dQ'i14}), noting that $(\Delta F)_{\lambda_{B}} = 0$ [or we used directly Crooks relation~(\ref{eq:Crooks2})], to obtain
\begin{equation}
\label{eq:dQi9}
 \sum_{i=1,\lambda_{B}}^{N}P_{i}(\xi_{A})_{\lambda_{B}} e^{\beta\left(Q_{i} \right)_{\lambda_{B}}}= e^{\frac{(\Delta \overline{S})_{\lambda_{B}}}{ k_{B}}}.
\end{equation}
For the combined process of step 1 and 2, we obtain $(\Delta \overline{S})_{\xi_{A}}+(\Delta \overline{S})_{\lambda_{B}}=\Delta (\overline{S})^{\mathrm{eq}}$ since $\overline{S}$ is a state function. Defining the random heat for the overall process as the sum of the heat exchanged in steps 1 and 2, $Q=\left(Q\right)_{\xi_{A}}+\left(Q\right)_{\lambda_{B}}$, the total average over the full  phase space starting from state $\{A\}$ with the probability $P_{i}(A)$ is obtained from the product of Eqs.~(\ref{eq:dQi6}) and (\ref{eq:dQi9})
\begin{align}
\label{eq:dQi10}
&\sum_{i=1,\lambda_{B}}^{N}P_{i}(\xi_{A})_{\lambda_{B}} e^{\beta\left(Q_{i} \right)_{\lambda_{B}}} \sum_{i=1,\xi_{A}}^{N} P_{i}(A) e^{\beta\left(Q_{i} \right)_{\xi_{A}}} \nonumber \\
 &=\sum_{i=1,\lambda_{B}}^{N}P_{i}(\xi_{A})_{\lambda_{B}} \sum_{i=1,\xi_{A}}^{N}P_{i}(A) e^{\beta\left( \left(Q_{i} \right)_{\xi_{A}}+ \left(Q_{i} \right)_{\lambda_{B}}\right)} \nonumber \\
 &= \sum_{i=1,\lambda_{B}}^{N}P_{i}(\xi_{A})_{\lambda_{B}} \sum_{i=1,\xi_{A}}^{N}P_{i}(A) e^{\beta Q_{i}} \nonumber \\
 &= \overline{e^{\beta Q}}=e^{\frac{\Delta (\overline{S})^{\mathrm{eq}}}{ k_{B}}}.
\end{align}
This is the nonequilibrium heat relation, which is equivalent to the nonequilibrium work relation but formulated for the random heat transferred to the thermal bath during the overall transformation between the two equilibrium states $\{A\}$ and $\{B\}$. A related expression for the entropy difference has been obtained by Adib for large systems undergoing isoenergetic processes, where $U_{B}=U_{A}$ and thus $W=Q$ in this case \cite{ADIB}.

\subsection{Summary and discussion}
\subsubsection{Thermodynamic interpretation of nonequilibrium relations}    
The nonequilibrium work relation implies that, from a series of out-of-equilibrium experiments involving random work, one can obtain the minimal work required to drive the system from an initial equilibrium state $\{A\}$ to a final equilibrium state $\{B\}$ along a quasistatic (equilibrium) path. This minimal work corresponds to the Helmholtz free-energy difference, $W^{\mathrm{min}} = \Delta F^{\mathrm{eq}}$. Similarly, the nonequilibrium heat relation indicates that, during the same process and from the same set of experiments, it is possible to access the minimal amount of heat exchanged with the thermal bath, which corresponds to the equilibrium entropy variation $Q^{\mathrm{min}} = \Delta (\overline{S})^{\mathrm{eq}} / \beta k_{B}$. Minimal work is thus associated with minimal heat in a way that ensures the first law of thermodynamics is satisfied. In this case, heat is transferred to the bath when positive work is performed on the system. Outside equilibrium, the average work is greater than $\Delta F^{\mathrm{eq}}$, whereas the magnitude of the average heat transferred to the bath is greater than $\left|\Delta (\overline{S})^{\mathrm{eq}}\right| / \beta k_{B}$. In the reverse transformation from $\{B\}$ to $\{A\}$, both the work delivered by the system to the surroundings along an equilibrium path and the heat absorbed by the system from the bath are maximal. In any case their sum remains equal to the equilibrium internal energy difference $\Delta U^{\mathrm{eq}}$ for the $A \rightarrow B$ transformation, and equal to $-\Delta U^{\mathrm{eq}}$ for the $B \rightarrow A$ transformation.
 
\subsubsection{Links between all the random variables}  
Figure~5 provides a schematic summary of the nonequilibrium relations discussed previously. 
\begin{figure}[!htbp]
  \centering
  \includegraphics[width=8.6 cm]{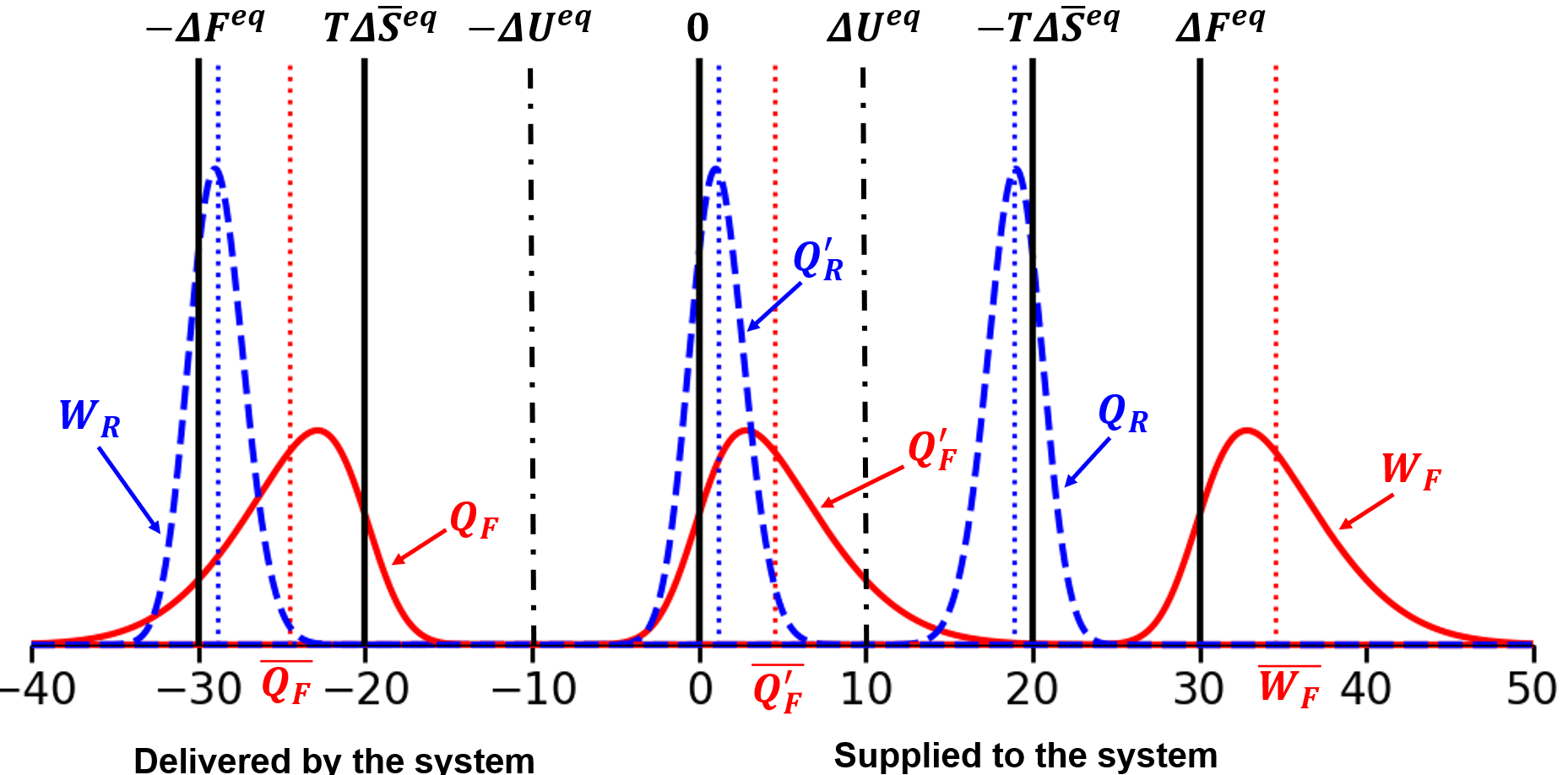}
  \caption{The fluctuations of work and uncompensated heat of Clausius as well as that of heat are shown in the same graph, with the red solid line for the forward transformation and the blue dashed line for the reverse transformation, each spread around their respective mean values. The abscisse represents energy in arbitrary units of Joule. The quantities above the zero axis are positive and supplied to the system, whereas those below are negative and delivered by the system to the surroundings.}
  \label{fig:Fig5}
\end{figure}
It illustrates the fluctuations in work, uncompensated heat of Clausius (as in Fig.~4), and heat, all shown as red solid lines. It also illustrates the fluctuations corresponding to the reverse transformation from $\{B\}$ to $\{A\}$ (in blue dashed lines). In this schematic representation, we assume that the equilibrium internal energy change $\Delta U^{\mathrm{eq}}$ between the states $\{A\}$ and $\{B\}$ is positive. If the small system under consideration were a small volume of an ideal gas (e.g. in Fig.~1), this internal energy variation would vanish since only isothermal transformations are considered. The horizontal axis represents energy in arbitrary units of Joules (a.u.). The value of $\Delta U^{\mathrm{eq}}$ is 10 a.u., that of $\Delta F^{\mathrm{eq}}$ is 30 a.u., and $T\Delta (\overline{S})^{\mathrm{eq}}$ takes the value of $-$20 a.u., consistent with the thermodynamic identity $\Delta F^{\mathrm{eq}}=\Delta U^{\mathrm{eq}}-T\Delta (\overline{S})^{\mathrm{eq}}$. In this context, $T\Delta (\overline{S})^{\mathrm{eq}}<0$. Likewise, the average work $\overline{W}$ lies 4.5 a.u. above $\Delta F^{\mathrm{eq}}$, and the average heat $\overline{Q}$ lies about 4.5 a.u. below $T\Delta (\overline{S})^{\mathrm{eq}}$, in agreement with the two equivalent formulations of the second law of thermodynamics $\overline{W}=\Delta F^{\mathrm{eq}}+\overline{Q'}$ and $\overline{Q}=T\Delta (\overline{S})^{\mathrm{eq}}-\overline{Q'}$. This follows from the fundamental statement of the second law $\overline{Q'}>0$, with $\overline{Q'}$ taking a value of 4.5 a.u. in our scheme. Only mean values for the $\{A\}$-to-$\{B\}$ transformation are shown below the red dotted lines. The thermodynamic identity $\Delta U^{\mathrm{eq}}= \overline{W}+\overline{Q}$, corresponding to the first law, is also satisfied. All associated fluctuation curves are shown as red solid lines around these mean values for the $\{A\}$-to-$\{B\}$ transformation. The fluctuation curves of work and uncompensated heat are identical in shape, although they are spread around different means, as demonstrated in the previous sections. This becomes particularly evident if we consider a special transformation with $\Delta F^{\mathrm{eq}}=0$, for which the two curves coincide and $\Delta U^{\mathrm{eq}}=T\Delta (\overline{S})^{\mathrm{eq}}$ holds. In the figure, the work and heat fluctuation curves are mirror images of each other, symmetric with respect to the axis at $\Delta U^{\mathrm{eq}}/2$. This situation is further illustrated by a purely entropic transformation with $\Delta U^{\mathrm{eq}}=0$ or $\Delta F^{\mathrm{eq}}=-T\Delta (\overline{S})^{\mathrm{eq}}$, where $\overline{W}=-\overline{Q}$. Similarly, the fluctuation curves of random heat and uncompensated heat are mirror images of each other, symmetric with respect to the axis at $T\Delta (\overline{S})^{\mathrm{eq}}/2$. This is evident if we consider a purely energetic (mechanical) transformation for which $T\Delta (\overline{S})^{\mathrm{eq}}=0$ or $\Delta F^{\mathrm{eq}}=\Delta U^{\mathrm{eq}}$, and in that case $\overline{Q}=-\overline{Q'}$. For the reverse transformation starting from the equilibrium state $\{B\}$ and reaching the equilibrium state $\{A\}$ with $\lambda$ varying over the same time interval $\tau $ with $d\lambda/dt(B\rightarrow A)=-d\lambda/dt(A\rightarrow B)$, all equilibrium values change sign. The average work $\overline{W}$ becomes negative, bounded below in magnitude by the maximum work the system can deliver to the surroundings $\left|\overline{W}\right|<\left|\Delta F^{\mathrm{eq}}_{B/A}\right|=\Delta F^{\mathrm{eq}}_{A/B}$, while the average heat $\overline{Q}$ becomes positive, bounded above by the maximum heat that the system can absorb from the environment $\overline{Q}<T\Delta (\overline{S})^{\mathrm{eq}}_{B/A}=-T\Delta (\overline{S})^{\mathrm{eq}}_{A/B}$. However, for this reverse transformation, nothing can be said about the precise shape of the fluctuation curves, which generally differ from those of the forward transformation from $\{A\}$ to $\{B\}$. This is because the nonequilibrium processes are generally different in this case. The only information available is that $\overline{Q'}>0$ also holds for this reverse case. All fluctuation curves corresponding to the reverse transformation are depicted with blue dashed lines. In this case, we choose $\overline{Q'}=$ 1.1 a.u. One may expect that a value of the switching rate $-d\lambda/dt$ exists in the reverse transformation (and thus a corresponding switching time) for which the fluctuation curves are the same as in the forward transformation. In that case, the fluctuations of all random variables exhibit the same shape and amplitude in both the forward and reverse transformations. The forward and reverse fluctuation curves of uncompensated heat of Clausius coincide exactly.

\begin{acknowledgments}
The author thanks  E. Collin, H. Guillou, M. Gibert, O. Bourgeois, N. Paillet, Ph. Camus, S. Panayotis, N. Aubergier, B. Brisuda, E. Porcheron (Institut Néel, Grenoble, France) and M. Peyrard (ENS Lyon) for numerous discussions and judicious comments.
\end{acknowledgments}

\appendix
\section{Application to a gas enclosed in a vessel with a moving piston}
\label{A}
The textbook example considered in Fig.~1 is a quantity of gas (ideal or interacting particles) enclosed in a vessel equipped with a movable piston. This model system has previously been studied at small scales within the framework of stochastic thermodynamics, where fluctuations are relevant. Lua and Grosberg used the Maxwell--Boltzmann velocity distribution to evaluate the number of particle–piston collisions \cite{LUA}. From this they determined the work performed by the gas and calculated the average exponential of the work to derive the Jarzynski identity. Similarly, Jarzynski considered an ideal gas of noninteracting point particles confined in a box closed by a piston to investigate the difference between typical realizations and dominant realizations \cite{JAR4}. In Jarzynski's paper, figs.~3 and~4 provide a clear illustration of rare event fluctuations. Figure~3 in the Jarzynski paper addresses the situation corresponding to step 1 of our analysis [see~Fig~6(b)]. As in the previous reference, the work is obtained from the cumulative change in kinetic energy of the particles during their successive collisions with the piston. Crooks and Jarzynski considered the same model, extended to interacting particles, and derived the exact distribution of work values for a slowly varying work protocol \cite{JAR5}. A striking result of the studies by Crooks and Jarzynski is that, due to the adiabatic invariance of the perturbation, the distribution remains canonical throughout the entire evolution of the external parameter. Although the distribution is not Gaussian for small particle numbers, it remains canonical. Our approach differs in that the time dependence of the energy is incorporated directly into the distribution, even at fixed values of the external parameter, resulting in a distribution that is not canonical. In the previous works cited above, the nonequilibrium character of the process is introduced through the concept of dissipated work, defined as the excess of the performed work over the work that would be obtained in a reversible isothermal process. This quantity corresponds to the uncompensated heat of Clausius in our manuscript. In our formulation, this quantity is introduced directly at the microscopic scale through a perturbation of the nonequilibrium canonical distribution, in which the internal variable $\xi$ appears in the dependence of the energy of the accessible microstates of the nonequilibrium small system considered. In this way, our formalism applies both to slow and to rapid variations of the external parameter. 

To illustrate our point using this textbook system, the meaning of $\xi$ is straightforward: It corresponds to the volume of the system $\xi=V_{s}$. The variable $\lambda $ denotes the vessel volume $\lambda = V_{v} $, which is externally controlled by the experimentalist. We therefore distinguish the system volume, which is an internal variable, from the vessel volume, which is externally imposed. The distinction between internal and external variables in statistical physics is discussed in several monographs (e.g., Ref.\cite{BAZA}). Let us consider in Fig.~6 a volume expansion under isothermal conditions with $\lambda = V_{v}$ and $\xi=V_{s}$. The work protocol begins at $V_{v}(A)$ at $t=0$ and ends at $V_{v}(B)$ at $t=\tau$, following a prescribed rate $dV_{v}/dt$ of change of the vessel volume. 
\begin{figure}[h]
  \centering
  \includegraphics[width=8.6 cm]{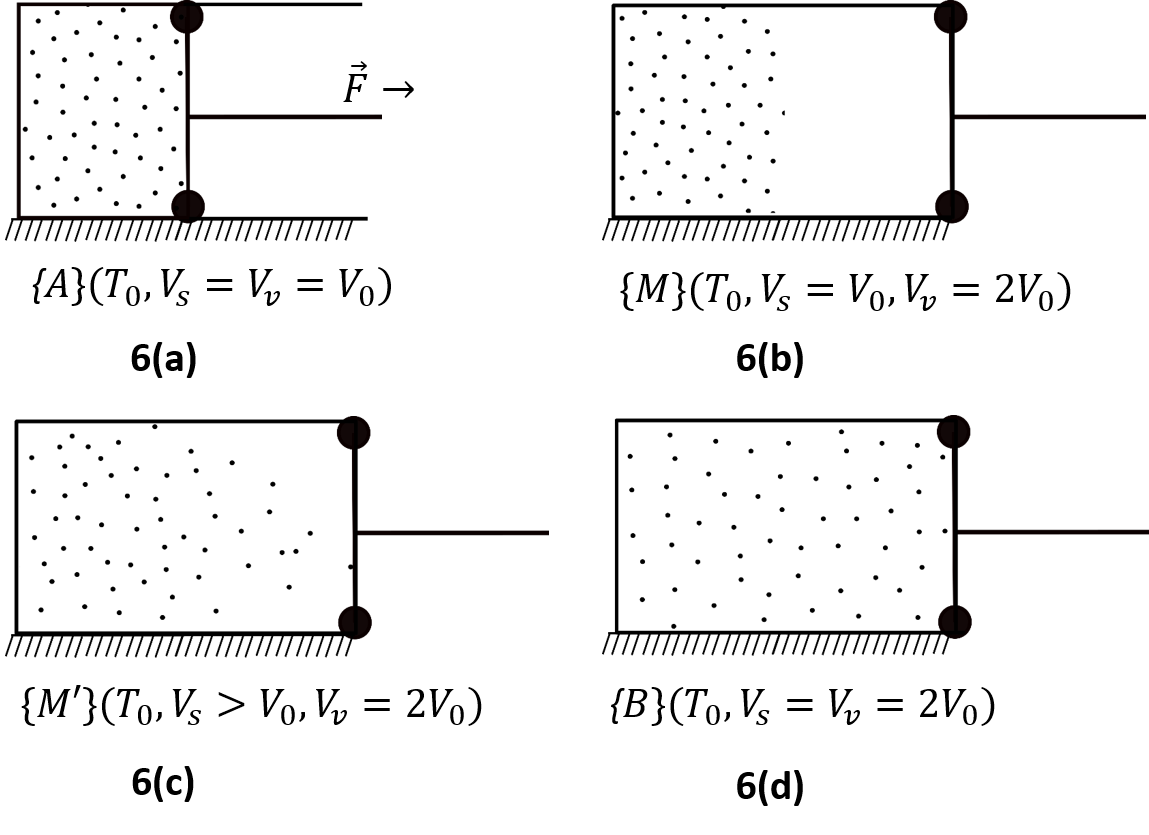}
  \caption{The model system in Fig.~1 is used to simulate the two-step work protocol described in the main text from an initial state $\{A\}$ to a final state $\{B\}$.
(a) Equilibrium state $\{A\}$: The system, with volume $V_{s}=V_{0}$, is enclosed in an external vessel of volume $V_{v}=V_{0}$.
(b) Nonequilibrium state $\{M\}$: The system volume remains fixed at $V_{s}=V_{0}$, while the external vessel volume is suddenly increased to $V_{v}=2V_{0}$.
(c) Nonequilibrium state $\{M'\}$: The system volume $V_{s}$ relaxes toward its equilibrium value corresponding to $V_{v}=2V_{0}$, producing entropy.
(d) Equilibrium state $\{B\}$: The system reaches $V_{s}=2V_{0}$ and is enclosed in an external vessel of volume $V_{v}=2V_{0}$.} 
  \label{fig:Fig6}
\end{figure}
If $\tau$ is large compared with the characteristic relaxation times of the system (associated with collisions, viscous transport, or acoustic propagation), then the transformation is quasistatic and the system remains in equilibrium at all times. In that case, the volume of the system is always equal to that of the container. In equilibrium, there is therefore no distinction between these two variables, and the internal pressure equals the external pressure exerted by the piston $P_{s}=P_{\mathrm{int}}=P_{\mathrm{ext}}$ (the external pressure is the force per unit area exerted by the piston). If, however, the vessel volume changes on a timescale shorter than the relaxation time of the particles, the system volume and the vessel volume differ [see Figs.~6(b) and~6(c)]. Only a small fraction of particles (those with the highest velocities) interact with the piston wall, and the internal pressure differs from the external pressure. In the limit of an infinitely rapid change of the vessel volume, the system volume remains effectively frozen at its initial value $V_{s}(A)$. It is in a nonequilibrium frozen state $\{M\}$ [see Fig~6(b)]. This corresponds to step 1 of the process discussed in the manuscript and to fig.~3(a) in Jarzynski’s paper. After this sudden change we have $V_{v}=V_{v}(B)$ (or $\lambda=\lambda_{B}$) while $V_{s}(\tau)=V_{v}(A)$. The internal variable $\xi$ is thus frozen at its initial value. Since the piston is now fixed, no further work is performed, although the system has not yet reached its final equilibrium state. Step 2 corresponds to the subsequent relaxation process. The dynamics of the internal variable $\xi$ depends on the microscopic properties of the gas (ideal gas, interacting particles, etc.). As a first approximation, one may describe the relaxation of the system volume by
\begin{equation}
\label{eq:rate}
\frac{d\xi}{dt}=\frac{dV_{s}}{dt}=-\frac{(V_{s}-V_{v})}{\tau_{\mathrm{relax}}},
\end{equation}
where $\tau_{\mathrm{relax}}$ is the characteristic relaxation time associated with the irreversible expansion of the gas in a larger volume. The entropy production associated with this relaxation process is given by the product of the thermodynamic force and the volume change
\begin{equation}
\label{eq:sigmai}
\delta_{i}S=\frac{(P_{\mathrm{ext}}-P_{\mathrm{int}})}{T} dV_{s}.
\end{equation}
The corresponding uncompensated heat of Clausius is
\begin{equation}
\label{eq:uncompens}
\delta Q'=(P_{\mathrm{ext}}-P_{\mathrm{int}}) dV_{s}.
\end{equation}
When the system is “small” and considered at the microscopic scale, fluctuations (pressure fluctuations of the system) render these quantities stochastic. If the work protocol starts from a microscopic state $i$ among $N$ possible states, the uncompensated heat becomes
\begin{equation}
\label{eq:uncompensi}
\delta Q'_{i}=[P_{\mathrm{ext}}-(P_{\mathrm{int}})_{i}] dV_{s}.
\end{equation}
The corresponding microscopic affinity for state $i$ is
\begin{equation}
\label{eq:affinityi}
A_{i}=[P_{\mathrm{ext}}-(P_{\mathrm{int}})_{i}].
\end{equation}
Fluctuations of the gas pressure on the piston wall therefore generate fluctuations of entropy production, which in turn lead to fluctuations of work and heat. For a gas expansion $\Delta V_{v}>0$, the average pressure difference is positive $(P_{\mathrm{ext}}-P_{\mathrm{int}})>0$, so that the average entropy production is positive. At the microscopic level, however, rare fluctuations may produce realizations for which this difference becomes negative, leading to negative entropy production for a single trajectory. Physically, this corresponds to situations in which the vessel expands faster than the typical particle velocity, but rare fluctuations in the tail of the velocity distribution produce particles energetic enough to collide with the piston and generate an internal pressure larger than the external one. This mechanism is illustrated in the manuscript using the simplified three–energy–level system in Fig.~3. More generally, for any irreversible process one can always introduce an internal variable that differs from the externally controlled parameter when the system is out of equilibrium. Our formulation simply expresses this statement at the mesoscopic scale, where fluctuations become significant.

\section{Derivation of the extended Gibbs canonical distribution}
\label{B}
Let us consider the macroscopic composite system ($S$) shown in Fig.~7, consisting of a small system ($s$) in perfect thermal contact with a large thermal reservoir ($R$). 
\begin{figure}[h]
  \centering
  \includegraphics[width=8.6 cm]{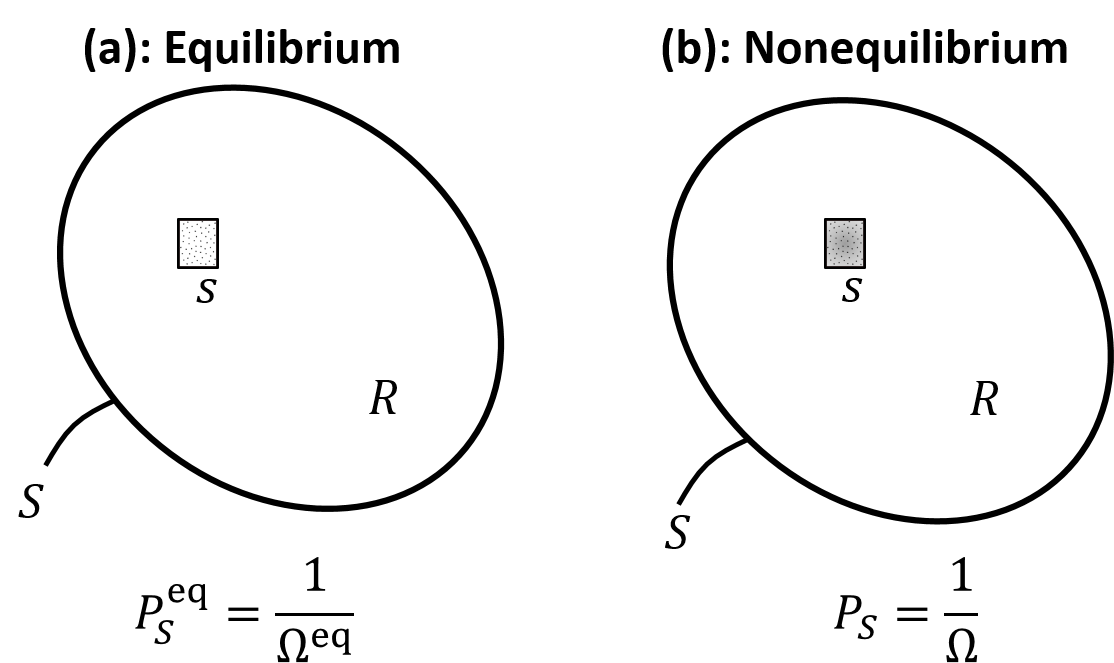}
  \caption{A macroscopic composite system $(S)$ of constant total energy $U_{S}$, consisting of a small system $(s)$ in perfect thermal contact with a large thermal reservoir $(R)$. (a) The small system is at equilibrium. All accessible microstates of the isolated composite system are equally probable, each with probability $1/\Omega^{\mathrm{eq}}$, where $\Omega^{\mathrm{eq}}$ is the constant total number of accessible microstates. (b) The small system is out of equilibrium, and consequently so is the composite system. Since the total energy remains constant, all accessible microstates of the isolated composite system remain equally probable but the total number of accessible microstates is modified and becomes $\Omega$. } 
  \label{fig:Fig7}
\end{figure}
The composite system is assumed to be isolated, and the interaction energy between ($s$) and ($R$) is neglected. The total energy is therefore conserved:
\begin{equation}
\label{eq:Conserv}
U_{S}=U_{R}+U_{s}=\mathrm{constant}.
\end{equation}
At statistical equilibrium, the small system occupies one of its microscopic states of energy $E_s^{\mathrm{eq}}$ according to the Gibbs canonical distribution. In the main text, we extended this description to situations where irreversible processes occur inside the boundaries of the small system, so that it may be driven out of equilibrium. The energy of such a nonequilibrium state is denoted $E_s$. Throughout the main text, we assumed $E_{s}^{\mathrm{eq}}=E_{s}^{\mathrm{eq}} (\lambda)$, whereas $E_{s}=E_{s} (\lambda,\xi)$, with $\lambda$ an external control parameter and $\xi$ an internal macroscopic variable characterizing the nonequilibrium state. Our objective in this Appendix is to determine the probability that the small system occupies a nonequilibrium state of energy $E_{s}$ and to show that it is given by Eq.~(\ref{eq:Pi0}). When the small system is fixed in a given microscopic state, either at equilibrium with an energy $E_{s}^{\mathrm{eq}}$ or out of equilibrium with an energy $E_{s}$, the number of states of the composite isolated system is entirely determined by the number of state of the reservoir $g_{R+s}=g_{R} g_{s}=g_{R}$ since the small system occupies one single specified state ($g_{s}=1$). The reservoir therefore has $g_{R }(U_{S}-E_{s}^{\mathrm{eq}})$ accessible states when the small system is at equilibrium and $g_{R }(U_{S}-E_{s})$ accessible states when it is out of equilibrium. According to the fundamental postulate of the microcanonical ensemble, all accessible states of the isolated composite system are equally probable. Let $\Omega^{\mathrm{eq}}$ denote the total number of accessible states of the composite system when the small system is at equilibrium and $\Omega$ the corresponding number when it is out of equilibrium. Consequently, the ratio of probabilities of observing the small system in an equilibrium and nonequilibrium states is
\begin{equation}
\label{eq:Ratio1}
\frac{P(E_{s}^{\mathrm{eq}})}{P(E_{s})}=\frac{\frac{g_{R }(U_{S}-E_{s}^{\mathrm{eq}})}{\Omega^{\mathrm{eq}}}}{\frac{g_{R }(U_{S}-E_{s})}{\Omega}}=\frac{\Omega}{\Omega^{\mathrm{eq}}}\frac{g_{R }(U_{S}-E_{s}^{\mathrm{eq}})}{g_{R }(U_{S}-E_{s})}.
\end{equation}
Introducing the microcanonical entropy of the reservoir $\sigma_{R}= \ln g_{R}$, Eq.~(\ref{eq:Ratio1}) becomes
\begin{equation}
\label{eq:Ratio2}
\frac{P(E_{s}^{\mathrm{eq}})}{P(E_{s})}=\frac{\Omega}{\Omega^{\mathrm{eq}}} \frac{e^{\sigma_{R}(U_{S}-E_{s}^{\mathrm{eq}})}}{e^{\sigma_{R}(U_{S}-E_{s})}} .
\end{equation}
Because the reservoir is macroscopically large, its entropy may be expanded to first order around $U_{S}$ in the two cases
\begin{subequations}
\begin{equation}
  \label{eq:microentrop1}
 \sigma_{R}(U_{S}-E_{s}^{\mathrm{eq}})=\sigma_{R}(U_{S})-E_{s}^{\mathrm{eq}}\left( \partial \sigma_{R}/\partial U_{S} \right)_{\lambda},
\end{equation}
\begin{equation}
  \label{eq:microentrop2}
   \sigma_{R}(U_{S}-E_{s})=\sigma_{R}(U_{S})-E_{s}\left( \partial \sigma_{R}/\partial U_{S} \right)_{\lambda, \xi}.
\end{equation}
\end{subequations} 
Throughout this work we assumed that the small system remains in thermal equilibrium with the reservoir at every instant, even while it is internally out of equilibrium. Consequently, both derivatives are equal to the inverse thermal energy
\begin{equation}
\label{eq:Temp}
\beta=\left( \partial \sigma_{R}/\partial U_{S} \right)_{\lambda}=\left( \partial \sigma_{R}/\partial U_{S} \right)_{\lambda, \xi}=\frac{1}{k_{B}T}.
\end{equation}
Substituting Eqs.~(\ref{eq:microentrop1}),(\ref{eq:microentrop2}) and~ (\ref{eq:Temp}) into Eq.~(\ref{eq:Ratio2}) yields
\begin{equation}
\label{eq:Ratio3}
\frac{P(E_{s}^{\mathrm{eq}})}{P(E_{s})}=\frac{\Omega}{\Omega^{\mathrm{eq}}}\frac{e^{-\beta E_{s}^{\mathrm{eq}}}}{e^{-\beta E_{s}}} .
\end{equation}
Since the equilibrium probability is given by the canonical distribution $P(E_{s}^{\mathrm{eq}} )=e^{-\beta E_{s}^{\mathrm{eq}}}/Z^{\mathrm{eq}}$, the probability of observing the nonequilibrium state immediately follows
\begin{equation}
\label{eq:probanoneq}
P(E_{s})=\frac{\Omega^{\mathrm{eq}}}{\Omega}\frac{e^{-\beta E_{s}}}{Z^{\mathrm{eq}}}.
\end{equation}
Finally, imposing the normalization condition on the nonequilibrium probabilities $\sum P(E_{s})=1 $ gives
\begin{equation}
\label{eq:Znoneq}
Z=\frac{\Omega}{\Omega^{\mathrm{eq}}}Z^{\mathrm{eq}},
\end{equation}
which is the generalized partition function of the nonequilibrium system. Eq.~(\ref{eq:probanoneq}) is therefore the extended canonical probability distribution for a nonequilibrium small system. In the main text, it is written in the generalized form of Eq.~(\ref{eq:Pi0}), where $E_{s}=E_{s} (\lambda,\xi)$ and $Z=Z(\beta,\lambda,\xi)$.

\section{$\xi$ as a state variable for the system outside equilibrium}
\label{C}
The macroscopic state $\{M\}$ of a system out of equilibrium can be characterized by at least three independent variables: the temperature $T$; a variable $\lambda$, which accounts for work exchange with the surroundings; and a variable $\xi$, which represents the internal disequilibrium of the system \cite{PRIGO}. In general, $\xi$ is an extensive variable describing the distribution of matter within the system (for instance, in the case of a chemical reaction, $\xi$ corresponds to the advancement of the reaction; in the example in Fig.~1 it corresponds to the volume of the gas) \cite{PRIGO}. The intensive variable conjugate to $\xi$ is the thermodynamic affinity $A$ (more precisely $A/T$). The total differential of the variable $\lambda$ can then be expressed as
\begin{equation}
\label{eq:AN1}
 d\lambda=\left(\frac{\partial \lambda}{\partial T}\right)_{A, \xi}dT+\left(\frac{\partial \lambda}{\partial A}\right)_{T, \xi}dA+\left(\frac{\partial \lambda}{\partial \xi}\right)_{T, A}d\xi.
\end{equation}
For an isothermal transformation, the state of the system depends on only two independent variables. In this case, one may write $\lambda = f(A,\xi)$. The corresponding total differential is then given by
\begin{equation}
\label{eq:AN2}
 d\lambda=\left(\frac{\partial \lambda}{\partial A}\right)_{\xi}dA+\left(\frac{\partial \lambda}{\partial \xi}\right)_{A}d\xi.
\end{equation}
It follows immediately that, for a transformation driving the system away from equilibrium, the variation of $\lambda$ and $\xi$ is accompanied by the generation of affinity. If $\xi$ is frozen in (kept constant throughout the transformation), the evolution of $\lambda$ is necessarily associated with a corresponding evolution of $A$. In contrast, for a transformation at equilibrium, the conditions $A=0$ and $dA=0$ hold simultaneously, and therefore
\begin{equation}
\label{eq:AN3}
 d\lambda=\left(\frac{\partial \lambda}{\partial \xi}\right)^{\mathrm{eq}}_{A=0}d\xi^{\mathrm{eq}}.
\end{equation}
At equilibrium, there exists a unique relation between $\lambda$ and $\xi^{\mathrm{eq}}$, implying that a single variable suffices to characterize the equilibrium state of the system. On the microscopic level, only conservative forces are present, and statistical equilibrium holds.

\end{document}